\documentclass[draft]{agujournal2019}
\usepackage{amsmath}
\usepackage{url} 
\usepackage{lineno}
\usepackage[inline]{trackchanges} 
\usepackage{soul}
\draftfalse

\journalname{JGR: Space Physics}

\begin{document}

%
%


\title{Global-MHD Simulations using \texttt{MagPIE} : Impact of Flux Transfer Events on the Ionosphere}

%
%




\authors{Arghyadeep Paul\affil{1}, Antoine Strugarek\affil{2}, Bhargav Vaidya\affil{1}}


\affiliation{1}{Department of Astronomy Astrophysics and Space Engineering, Indian Institute of Technology Indore,Khandwa Road, Simrol, Indore 453552, India}
\affiliation{2}{Université Paris-Saclay, Université Paris Cité, CEA, CNRS, AIM, 91191, Gif-sur-Yvette, France}




\correspondingauthor{Arghyadeep Paul}{arghyadeepp@gmail.com}




\begin{keypoints}
\item A coupled magnetosphere-ionosphere module named \texttt{MagPIE} is developed to study the ionospheric impact of Flux Transfer Events (FTEs).
\item Cusp-FTE reconnection has a significant impact on the ionospheric field aligned currents (FACs).
\item The imprint of an FTE on the ionosphere is generally characterised as a combination of an I-shaped patch surrounded by a U-shaped patch of FACs.
\item FTE signatures on the ionosphere resemble the morphology of discrete dayside auroral arcs.
\item FTEs are seen to produce vortex-like ionospheric flow patterns.
\end{keypoints}

%
%

%
%


\begin{abstract}
This study presents a recently developed two-way coupled magnetosphere-ionosphere model named ``\texttt{MagPIE}" that enables the investigation of the impact of flux transfer events (FTEs) on the ionosphere. Our findings highlight the prominent role of cusp-FTE reconnection in influencing the ionosphere. The typical morphology of an FTE signal, represented by field-aligned currents (FACs) on the ionosphere, is shown to exhibit a distinct pattern characterized by an `I'-shaped patch surrounded by a `U'-shaped patch. Furthermore, we demonstrate that the effects of FACs resulting from FTEs may extend well into the region of closed field lines on the ionosphere. These FACs are seen to exhibit a remarkable resemblance to discrete dayside auroral arcs, providing further evidence that FTEs can be considered as a probable cause of such phenomena. Additionally, FTEs generate vortex-like patterns of ionospheric flow, which can manifest as either twin vortices or a combination of multiple vortices, depending on the characteristics of the FACs producing them. Furthermore, we present compelling evidence of morphological similarity between the simulated ionospheric signatures obtained from the \texttt{MagPIE} model and an observation made by the SWARM satellites. The agreement between our model and observational data further strengthens the credibility of our model and opens up new avenues to theoretically explore the complex ionospheric effects caused by FTEs.
\end{abstract}

\section*{Plain Language Summary}
We introduce a new numerical model called ``MagPIE" that facilitates the understanding of how events in space called flux transfer events (FTEs) affect the Earth's ionosphere. The Earth's magnetosphere comprises of the northern and the southern polar cusps. The study shows that a particular type of interaction between FTEs and the polar cusps, called cusp-FTE reconnection, has a significant impact on the ionosphere. When cusp-FTE reconnection occurs, it imposes a specific imprint of plasma currents on the ionosphere that look like an 'I' shape surrounded by a 'U' shape. These currents have effects beyond the immediate area where the FTEs connect to the ionosphere. The shape of these currents in the ionosphere is similar to how discrete bright auroral arcs appear in the polar regions, which suggests that FTEs could be a cause of these auroral phenomena. It is seen that that these currents produced by ``\texttt{MagPIE}" model  are similar to an observation made by ionospheric satellites which strengthens the reliability of the model and provides new opportunities for exploring the complex effects of FTEs on the ionosphere.

\section{Introduction} \label{sec:intro}
 Interactions between the solar wind and the magnetosphere form the fundamental basis of space physics, magnetospheric physics, ionospheric physics, and space weather \cite{Wing_2023}. The magnetosphere and the ionosphere behave as an intricately dependent system of domains at the macroscopic as well as the microscopic scales.
 
 Field Aligned Currents (FACs) facilitate the transfer of energy and momentum between the magnetosphere and the ionosphere \cite{Luhr_2021}. Their existence was proposed by \citeA{birkeland_1908} and their effects were first detected by \citeA{Zmuda_1966} using data from polar orbiting satellites. \citeA{Iijima_1978} observed that for southward interplanetary magnetic field (IMF), the current systems form two concentric rings at ionospheric heights: a poleward ring (called the Region-1 or R1 current system) and an equatorward ring (called the Region-2 or R2 current system), both of which are powered by distinct regions of the magnetosphere. It is known that the R1 current system closes to itself across the polar cap and to the R2 currents in the auroral region with the help of Pedersen currents \cite{Iijima_1978, Coxon_2014}.  The closure of these Birkeland currents necessitates the formation of a potential difference across the polar cap which, in turn, gives rise to the characteristic twin-cell ionospheric convection pattern associated with the Dungey cycle \cite{Dungey_1961, Cowley_1992}. On the other hand, for northward IMF, it is seen that intense FAC patches appear poleward of the R1 FACs. These currents are commonly termed as R0 FACs and their morphology is known to strongly depend on the polarity of the IMF $B_Y$ \cite{Vennerstrom_2005,Luhr_2021}.

  The aforementioned connection between the magnetosphere and the ionosphere is heavily reliant on the solar wind-magnetosphere interaction. The strength and morphology of the R1 and R2 FACs play a vital role in the magnetosphere-ionosphere coupling whereas the state of energy sources in these current systems depend on the degree of coupling between the solar wind and the magnetosphere  \cite{Tanaka_1995}. One of the key mechanisms that govern the microphysical aspects of solar wind-magnetosphere interaction is the process of magnetic reconnection. It is known that reconnection at the Earth's magnetopause facilitates the inflow of mass, momentum and energy into the magnetospheric system. A correlation between geomagnetic activity and the southward directed interplanetary magnetic field (IMF) led \citeA{Dungey_1961} to suggest that magnetopause reconnection allows mass, momentum, and energy to enter the magnetosphere. However, whether reconnection is continuous or intermittent has long been debated since. Flux Transfer Events (FTEs) are a result of patchy and intermittent reconnection at the Earth's magnetopause \cite{Berchem_1984, Kawano_1996, Daum_2008}.

 The possibility that such patchy sporadic reconnection could occasionally be the dominant mode of momentum transfer between the solar-wind and the inner magnetosphere has boosted the perceived importance of understanding these transient phenomena \cite{Russell_1978,Rijnbeek_1984,Saunders_1984}. Their characteristics are typically a bipolar signature of the magnetic field component normal to the magnetopause surface along with an enhancement in the axial magnetic field component. This is also generally associated with a mixing of plasma populations of the magnetosheath and the magnetospheric kind \cite{Russell_1978, Daly_1981,Thomsen_1987}. These FTEs, over the course of their evolution, are expected to leave ionospheric traces. It has been postulated that the first signature of enhanced reconnection at the magnetopause propagates as an Alfv\'enic perturbation from the magnetopause to the ionosphere \cite{Glassmeier_1996}. Auroral features named Poleward Moving Auroral Forms (PMAFs) that brighten at the equatorward boundary and fade out as they move into the polar cap are widely being accepted as the ionospheric observational signatures of FTEs at the dayside magnetopause. Additional signatures include but are not limited to Flow Channel Events(FCEs) and Pulsed Ionospheric Flows also known as Poleward Moving Radar Auroral Forms \cite{Sandholt_1986,McWilliams_2000,Fear_2017}. \citeA{Oskavik_2004} characterised the enhanced flow channels corresponding to PMAFs using EISCAT Svalbard radar data. 

The ionospheric reaction to FTEs has been predicated by theoretical investigations during the last several decades. \citeA{Southwood_1987} gave a simplistic sketch of the current pattern in the form of an FAC pair that would be generated by a flux tube on the ionospehric surface. Later, \citeA{Southwood_1989} suggested that the same localised FAC pair, which is a miniaturised version of the polar cap, may be set up on the ionospheric surface in response to FTEs. They further imply that the coupling of the magnetosphere and the ionosphere may itself be involved in modulating the dayside reconnection rate. \citeA{Glassmeier_1996} suggested that the impulsive ionospheric flow bursts observed by \citeA{Lockwood_1988}  correspond well with the theoretical models proposed by \citeA{Southwood_1989}. The recurrence of the flow burst events also concur well with the recurrence rate of FTEs at the dayside magnetopause. However, they infer that a certain degree of ambiguity remains in defining the ionospheric and ground magnetic signatures of FTEs. \citeA{Crooker_1990} pointed out that the idealised round shape of existing FTE models on the ionosphere are unrealistic and signatures of FTEs must have a highly distorted ionospheric footprint. The ionospheric footprints they propose is found to resemble the fan shaped mid day auroral arc observed by \citeA{Akasofu_1976}. They therefore conclude that midday auroral arcs could be the ionospheric signatures of FTEs. \citeA{Kaufmann_1990} has also performed an analytical study of the distortions introduced while mapping magnetopause structures along magnetic field lines to the ionospheric surface.

 Ground-based equipments such as magnetometers, radars, and imagers routinely monitor transient phenomena in the ionosphere. However, it remains uncertain which perturbations in the ionosphere are associated with which specific magnetospheric event. The ambiguity is enhanced by the fact that simultaneous global observations of the combined magnetosphere-ionosphere system are scarce due to limited field of view of ground based instruments and in-situ probes. Moreover, large uncertainties exist in the mapping of the Earth's magnetic field lines from the magnetospheric transients to the ionosphhere that makes it  challenging to accurately trace events along magnetic field lines to the polar ionosphere. A statistical study by \citeA{Neudegg_2000} showed a correlation between FTEs at the magnetopause and their PMAFs. However, PMAFs have been recorded to form during both southward and northward IMF conditions. This is in contrast to subsolar FTEs which are dominantly observed for southward IMF leading to the inference that PMAFs and FTEs may not have a consistent correlation \cite{Fasel_1995}.  Simultaneous observation of FTEs and their ionospheric signatures has remained scarce. \citeA{Elphic_1990} associated an FTE observed by the ISEE2 satellites at the magnetopause to its subsequent ionospheric signature in the form of ionospheric flow bursts. \citeA{Marchaudon_2004} used simultaneous observations from the in-situ CLUSTER data at the cusp and the SuperDARN radars in the ionosphere to deduce the correlation between three FTEs and their associated particle injection events in the ionosphere. 

 Modelling studies on the effect of FTEs on the combined magnetosphere-ionosphere have also remained sparse. \citeA{Daum_2008} performed global magnetohydrodynamic simulations using the Block-Adaptive-Tree-Solarwind-Roe-Upwind-Scheme (BATS-R-US) code to trace FTE magnetic field lines at the magnetopause down to their ionospheric footpoints and partially reproduced the mapping distortions proposed by \citeA{Crooker_1990}. \citeA{Omidi_2007} leveraged 2.5-D global hybrid simulations to reveal that the FTEs produce bow waves that are responsible for plasma injection into the cusp and subsequent ionospheric signatures. They further highlighted that the interaction of the FTEs with the polar cusps lead to localised magnetic reconnection between the cusp field lines and FTE field lines. \citeA{Omidi_2007} also showed that such FTEs are also associated with a downward flux of energetic ions that have characteristics similar to PMAFs. Recently, \citeA{Grandin_2020} had presented a comparison on the auroral proton precipitation at the polar cusps during northward and southward IMF conditions using the hybrid Vlasov code VLASIATOR \cite{Alfthan_2014}. Their results highlight that bursty proton precipitation during southward IMF is associated with the transit of FTEs through the vicinity of the cusp.

 Despite the extensive research interest for decades, severe gaps still remain in the current understanding of the significance of FTEs in a combined magnetosphere-ionosphere scenario. At present, there is no direct means to observationally monitor the propagation of the FTE signal from the FTE flux rope to the ionosphere. Opportune simultaneous observations therefore rely on a possible delay of the FTE signatures at the high altitude cusp and their theoretically motivated and complementary observational signatures at the ionospheric surface to correlate FTEs with their low altitude signatures. The signatures of FTEs on the ionospheric surface in terms of the FACs and the ionospheric potential are also poorly understood at present. It is not yet clear if the simplistic model postulated by \citeA{Southwood_1989} would continue to hold true in a magnetosphere that is representative of the realistic scenario. It is also not clear what function the polar cusps play in the interaction between the FTE and the ionosphere and a lack of comprehension exists in the injection mechanism of perturbations at the cusps. This study presents, to the best of our knowledge, the first comprehensive assessment of the effects of FTEs on the global field and flow patterns in a two-way coupled global magnetosphere-ionosphere model. The global MHD magnetosphere facilitates the formation and evolution of FTEs and the coupled ionospheric model is leveraged to provide an outlook of the ionospheric response to these transient events. 
  
 The article is organised as follows: Section \ref{sec:numerical_framework} describes the model setup, the numerical framework along with the initial conditions. Section \ref{sec:iono_on_vs_off} explains the macroscopic differences upon the inclusion of an ionospheric model in the global MHD simulation. Section \ref{sec:fte_iono_interaction} and \ref{sec:sat_signatures} are dedicated towards the salient results corresponding to the FTE-ionosphere interaction. Finally, section \ref{sec:disc_and_summ} summarises the article and presents relevant discussions.

\section{Numerical Framework} \label{sec:numerical_framework}
\subsection{Global Magnetohydrodynamic Model}
    We describe here, our global magnetohydrodynamic model to study the interaction of the ambient solar wind with an Earth-like planetary magnetosphere. The core model has been developed by \citeA{Paul_2022} using the resistive-MHD module of the open source astrophysical gasdynamics code PLUTO \cite{Mignone_2007}. An additional ionospheric module has been added in the current work. The combined magnetosphere-ionosphere model has been named as \texttt{MagPIE} (\textit{\textbf{Mag}netosphere of \textbf{P}lanets with \textbf{I}onospheric \textbf{E}lectrostatics}). The model solves a set of conservation laws in 3 dimensions in the form of single-fluid MHD equations given as:
\begin{equation}\label{eq:MHD_eqs}
\begin{aligned}
    \frac{\partial \rho}{\partial t} + \nabla \cdot(\rho \mathbf{v})  &=  0 \\
    \frac{\partial (\rho  \mathbf{v})}{\partial t} + \nabla \cdot \left[\rho \mathbf{v}\mathbf{v} - \mathbf{B}\mathbf{B}\right] + \nabla \left( p + \frac{\mathbf{B}^2}{2} \right)  &= 0 \\
    \frac{\partial \mathbf{B}}{\partial t} + \nabla \times (c\mathbf{E}) &= 0 \\
    \frac{\partial E_{t}}{\partial t} + \nabla \cdot \left[ \left( \frac{\rho\mathbf{v}^2}{2} +\rho e + p\right)\mathbf{v}  + c\mathbf{E}\times \mathbf{B} \right] &= 0
\end{aligned}
\end{equation}
where $\rho$ is the mass density, $\mathbf{v}$ is the gas velocity, $p$ is the thermal pressure and $\mathbf{B}$ is the magnetic field. A factor of 1/$\sqrt{4\pi}$ has been absorbed in the definition of $\mathbf{B}$. $E_{t}$ is the total energy density which can be described as:
\begin{equation}\label{eq:energy}
    E_{t} = \rho e + \frac{\rho \mathbf{v}^{2}}{2} + \frac{\mathbf{B}^2}{2}
\end{equation}

An ideal equation of state provides the closure as $\rho e = p/ (\gamma -1)$ wherein, $\gamma$ is the ratio of specific heats having a value of 5/3. The electric field, \textbf{E}, is composed of a convective and a resistive component and is defined as:
\begin{equation}\label{eq:induction_eq}
    c\mathbf{E} = -\mathbf{v}\times \mathbf{B} + \frac{\eta}{c} \mathbf{J}
\end{equation}
where $\eta$ is the resistivity and $\mathbf{J}= c\nabla\times\mathbf{B}$ is the current density. In CGS units, the expression includes `c', which denotes the speed of light. To detach the dependence of the reconnection process on the numerical resistivity of the system, we incorporate an explicit  current density dependent resistivity model in the present study which is defined as: 
\[
    \eta = 
\begin{cases}
    1.016 \times 10^{9} \; (m^2/s),& \text{if } |\mathbf{J}|\geq \mathbf{J}_{\rm threshold}\\
    0.0 \;(m^2/s),              & \text{otherwise}
\end{cases}
\]

The physical effect of the diffusion coefficient is to broaden the 
 magnetopause current layer \cite{Komar_thesis}. The above choice of the diffusion coefficient ensures that the magnetopause current sheet is consistently resolved by at least 7-8 grid cells \cite{Paul_2022}. Additionally, this choice of the magnitude of the diffusion coefficient is also partially motivated by the fact that it sets the Lundquist number of the magnetopause current sheet at $\sim 6.7 \times 10^{4}$ (measured over a meridional slice with the Alfv\'en speed averaged over the magnetosheath and the magnetospheric values) which is larger than the threshold of occurrence of fast magnetic reconnection \cite{Bhattacharjee_2009}. Drawing from empirical insights, a current sheet spanning only 2-3 grid cells would imply that the explicit resistivity falls short of prevailing over the system's numerical resistivity. Conversely, a much higher resistivity would diffuse the current sheet even more, which would impact FTE formation. Hence, the magnitude of resistivity is optimized for the aims of this study. A careful choice of $\mathbf{J}_{\rm threshold} = 1.7 \times 10^{-8}Am^{-2}$ ensures that the resistivity only triggers reconnection at the magnetopause boundary layer and small regions of the high altitude cusp where $|\mathbf{J}| > \mathbf{J}_{\rm threshold}$. All other regions follow an ideal-MHD evolution. The resistivity is treated within the code using a second order accurate explicit multistage Runge-Kutta Legendre time stepping scheme 
 \cite{Vaidya_2017}. 

The flux computations have also been performed with the second order accurate Harten-Lax-vanLeer (HLL) solver and the solenoidal constraint ($\nabla\cdot\textbf{B}$ = 0) is imposed by coupling the induction equation to a  Generalised Lagrange Multiplier (GLM) and solving a modified set of conservation laws in a cell-centered approach \cite{Dedner_2002}. 

Additionally, the present formalism treats the total magnetic field $\mathbf{B}$ inside the domain in a split configuration given as 
\begin{equation}\label{eqn:split_field}
    \mathbf{B}(x,y,z,t) = \mathbf{B}_{0}(x,y,z) + \mathbf{B}_{1}(x,y,z,t)  
\end{equation}
where $\textbf{B}_0$ is a curl free, time invariant, background magnetic field and $\mathbf{B}_1$ behaves as a deviation. For such a configuration, the energy depends only on the deviation $\mathbf{B}_{1}$ which turns out to be computationally convenient when dealing with low-$\beta$ plasma. 

In recent times, beyond-MHD models have been developed to investigate magnetic reconnection at the dayside magnetosphere. These models incorporate kinetic physics into small-scale magnetospheric processes. Examples of such codes include the MHD-EPIC code by \citeA{Chen_2020} and the MHD-AEPIC code by \citeA{Wang_2022}, both of which integrate a localized particle-in-cell domain within a larger-scale MHD magnetospheric model. Additionally, hybrid codes have been devised that treat ions kinetically, with electrons serving as a charge-neutralizing background fluid \cite{Lin_2005, Omelchenko_2021}. Furthermore, advanced hybrid-Vlasov codes like VLASIATOR, evolve particle distribution functions to accurately capture kinetic effects in the magnetosphere \cite{Palmroth_2018,Palmroth_2022}. Despite the development of such a rich physics-based approach to depict magnetospheric processes more fundamentally, global-MHD models remain attractive for numerical magnetospheric studies owing to their high stability, relatively low computational cost, and the increasing challenge in obtaining sufficient computational resources. While resistive-MHD simulations may not accurately capture the intricacies of the reconnection diffusion region in a collisionless regime, they can effectively reproduce the macroscopic in-situ observational features and qualitative signatures of magnetic reconnection (FTEs, magnetic field signatures and flow signatures), when focusing on length scales greater than the ion gyroradius \cite{Rosenqvist_2008,Dorelli_2009}. Consequently, MHD simulations continue to be extensively employed and actively refined \cite{Sorathia_2017}. Additionally, global-MHD codes, operating at the limits of the MHD approximation, have recently been utilized to deduce the auroral signatures stemming from magnetospheric field-aligned currents \cite{Sorathia_2020}. Hence, while not at the cutting edge of numerical plasma simulations, the proposed framework is anticipated to offer an initial basis for investigating the influence of FTEs and their associated currents on the ionosphere.

\subsection{Initial and Boundary Conditions}\label{sec:Initial_and_boundary}
The set of resistive-MHD equations given by equation \eqref{eq:MHD_eqs} are solved in a Cartesian domain enclosed by $\rm -50 R_{E}\leq X \leq 100 R_{E}$, $\rm -100 R_{E}\leq Y \leq 100 R_{E}$ and $\rm -100 R_{E}\leq Z \leq 100 R_{E}$, where $\rm R_E$ denotes the Earth radius. The computational grid is a combination of a uniformly spaced cubical grid that is padded on all sides by a stretched grid up to the domain boundaries. The uniform grid covers a span of $\rm -15 R_{E}\leq X \leq 15 R_{E}$, $\rm -15 R_{E}\leq Y \leq 15 R_{E}$ and $\rm -15 R_{E}\leq Z \leq 15 R_{E}$ and has a resolution of 512$\times$512$\times$512 grid cells allowing for a smallest grid size of $\Delta x = \Delta y = \Delta z =$ 0.05$\rm R_E$ in this region. The grid cell size as well as the extent of the computational grid can be easily scaled as per requirement in \texttt{MagPIE}. Beyond the uniformly spaced grid, the size of the $N^{th}$ stretched grid cell varies approximately as $\Delta x \times 1.05^{N}$ along the (-X) direction and $\Delta x \times 1.07^{N}$ along all the other directions where N varies from 0 to 64. The stretched grid region simply serves as a buffer to move the boundaries away in order to minimise any effects that the numerical domain boundaries might have on the region of interest in this study. In order to emulate the magnetic field of an Earth-like planet, $\mathbf{B_{0}}$, as given in equation \eqref{eqn:split_field} is prescribed as a magnetic dipole placed at the origin of the simulation box having an equatorial field strength of $3 \times 10^{-5}$ T at 1$\rm R_E$. The axis of the dipole is aligned perfectly with the Z-axis in this study.

 The entire domain is initially filled with a low density plasma having a number density of 1 proton/cm$^{3}$. Thereafter, a solar wind inflow is prescribed at the left X-boundary with an inflow speed of $v_{sw}=$ 400kms$^{-1}$, a number density of 5 protons/cm$^{3}$ and a thermal pressure of 7 $\times 10^{-3}$nPa. The remaining five boundaries are set to have Neumann boundary conditions which mandates that all MHD variables flow out freely across these boundaries. The setup is initialised with a northward interplanetary magnetic field (IMF) with components $B_{x(sw)}= 0$nT, $B_{y(sw)}= -2.5$nT and $B_{z(sw)}= 5$nT that washes out the static plasma in the domain and establishes a magnetosphere. The magnetic field components mentioned above are given in the Geocentric Solar Magnetospheric (GSM) coordinate system. The X and Z coordinates of the simulation domain align with those of the GSM coordinate system. However, in the simulation domain, the Y component is inverted relative to the GSM coordinate system, meaning that $\hat{y}_{\rm GSM}$ = $-\hat{y}_{\rm simulation}$. After driving the magnetosphere for over an hour in this configuration, the IMF is turned southward ($B_{z(sw)}= -5$nT) and the magnetosphere is allowed to adapt to this new IMF configuration. We denote the time when the southward-IMF hits the bow-shock as t = 0s and the global times mentioned in the manuscript henceforth follow this convention.

 \subsection{Ionospheric Model}
 The global-MHD domain in \texttt{MagPIE} also exhibits a spherical internal boundary (IB) set at radii r$\leq$ 3.5 $R_E$ which enables a two-way magnetosphere-ionosphere coupling. The MHD formulation represented by the set of equations \ref{eq:MHD_eqs} is not solved within this region. Instead, the field-aligned currents (FACs , $J{_\parallel}$) calculated at this surface are used to drive an ionospheric model of the kind first described in \citeA{Goodman1995}, and the output of the global-MHD driven ionospheric model is used to impose the drift velocities at the surface of the IB. The ionospheric model in \texttt{MagPIE} corresponds to a formalism that treats the ionosphere as an infinitesimally thin spherical surface with height integrated conductance profiles. The current continuity equation within this approximation is given as:
\begin{equation}\label{eq:ionosph_cur_cont}
    \nabla_{\perp}\cdot \begin{bmatrix} \Sigma_P/\cos^2 \varepsilon & -\Sigma_H /\cos \varepsilon \\ \Sigma_H/\cos \varepsilon & \Sigma_P \end{bmatrix} \cdot \nabla_{\perp} \psi = J_{\parallel}\cos\varepsilon 
\end{equation}
where $\nabla_{\perp}$ is the component of the differential operator on the spherical shell, $\Sigma_P$, $\Sigma_H$ and $\varepsilon$ are the height integrated Pedersen conductance, Hall conductance and the magnetic dip angle respectively (the mathematical formula for the dip angle is given in equation \ref{eq:NH_cose} of the appendix).  Equation \ref{eq:ionosph_cur_cont} is obtained under the approximations that (a) all currents are primarily field aligned within the spherical shell between the IB, and the ionosphere at 1.05 $R_E$, (b) the electric field in the ionosphere is described as a potential field and (c) the conductance along the field lines is infinite leading to $E_{\parallel}=0$ \cite{Lotko_2004, Merkin_lyon_2010}. The complete mathematical framework used to model the ionosphere and the two-way magnetosphere-ionosphere coupling has been elaborated in \ref{appendix1}. A thermal pressure of 4.8 $\times 10^{-3}$nPa and a density of 350 protons/cm$^{3}$ is prescribed inside the IB. The magnetic field is considered to be a pure dipole inside the IB and the velocities on the IB surface are imposed from the output of the ionospheric module as described in \ref{appendix1}. 

The ionospheric module in \texttt{MagPIE} for the northern and southern hemispheres are initialised as 2D spherical grid surfaces having extents of [0$^{\circ}$ - 33.2$^{\circ}$] in $\theta$ and [0$^{\circ}$ - 360$^{\circ}$] in $\phi$ for the northern hemisphere (NH) ($\theta$ represents colatitude) and [146.8$^{\circ}$ - 180$^{\circ}$] in $\theta$ and [0$^{\circ}$ - 360$^{\circ}$] in $\phi$ in the southern hemisphere (SH). The $\theta$ extent of the ionospheric module in the two hemispheres are regulated by the radius of the IB as described by equation \eqref{eq:thetaLLB}. This ionospheric domain in each hemisphere is divided into 64 $\times$ 256 uniformly spaced grid cells in $\theta$ and $\phi$. The height integrated Pedersen ($\Sigma_P$) and Hall ($\Sigma_H$) conductances are specified to have constant values of 7 and 12 Mhos respectively \cite{Coxon_2016}. The NH and the SH are solved as two separate surfaces within the ionospheric module, however, the solutions near $\theta$ = 90$^{\circ}$ in the form of velocities, when imposed at the IB as a result of the two way coupling, is verified to be continuous. The solution from the ionospheric module is computed and imposed on the global-MHD internal boundary at regular intervals of 15 seconds whereafter, it is held constant until the next ionospheric solution is calculated.

\section{Effect of Ionospheric Coupling in \texttt{\large{M\MakeLowercase{ag}PIE}}} \label{sec:iono_on_vs_off}
    A two-way magnetosphere-ionosphere coupling, such as the one incorporated in \texttt{MagPIE}, is expected to modify the global-MHD solution due to the feedback from the ionospheric module.  In this section, we briefly summarise some of the clearly discernible differences seen upon the inclusion of an MI coupling into a global-MHD model. In order to have a basis for comparison on how the magnetosphere responds to the two-way coupling on a global level, we have performed an analogous run where the MI coupling has been turned off (we name this setup C-0 henceforth) in addition to the run that incorporates an MI coupling (named C-1 henceforth). 
    
\begin{figure*}
    \centering
    \includegraphics[width=1.0\textwidth]{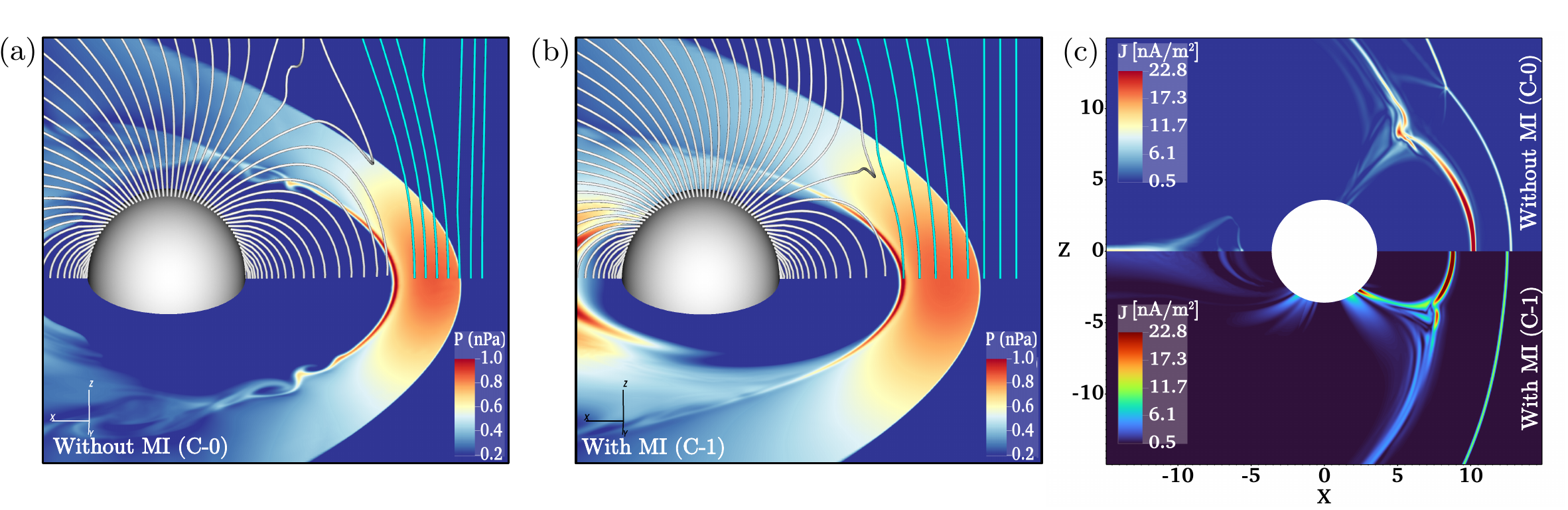}
    \caption{Panel (a) shows a zoomed-in portion of the global-MHD domain with the magnetosphere-ionosphere coupling turned off (setup named C-0), whereas panel (b) shows the MHD solution in the domain at the same time with the magnetosphere-ionosphere coupling turned on (setup named C-1). The background color in the plots represents thermal pressure. The white magnetic field lines in panels (a) and (b) represent the streamlines connected to the internal boundary surface (denoted by the white spherical surface) whereas the sky-blue field lines represent the IMF. The top half of panel (c) shows the X-Z slice of the current density ($\mathbf{J}$) magnitude in the domain corresponding to the C-0 setup, whereas the bottom half shows the same for the C-1 setup.}
    \label{fig:with_without_mi}
\end{figure*}

    Panel (a) of figure \ref{fig:with_without_mi} represents the C-0 setup at t = 4783s. Panel (b) is at the same time for the setup with the MI coupling enabled (C-1). In accordance with the theme of this paper, we mainly focus on the prominent differences on the dayside, the most salient of which is the steady-state magnetopause standoff distance. The subsolar magnetopause standoff distance(R$_{MP}$) for the C-0 setup was calculated to be $R_{MP}$ $\sim$ 10$R_E$. This measure is true if one regards the magnetopause standoff to purely be a balance between the Earth's magnetic pressure and the solar wind dynamic pressure omitting any consideration of the IMF direction. It is however observed that for steady southward IMF conditions, the magnetopause moves inwards by up to 1.5 $R_E$ and the phenomenon is termed as `magnetopause erosion' \cite{Pudovkin_1997}. As can be seen in panel (b) of figure \ref{fig:with_without_mi} the magnetopause indeed appears to be comparatively more eroded to a lower standoff distance. The standoff distance is found to be $R_{MP}$ $\sim$ 9$R_E$ for the C-1 setup. For a direct comparison, panel (c) of figure \ref{fig:with_without_mi} shows the magnitude of the current density $|J|$ for the two different setups. The top half of the image shows the C-0 setup and the bottom half shows the C-1 setup. As can be distinctively seen the magnetopause in the C-1 setup lies considerably inwards. Such an erosion of the magnetopause has been previously reported in observations \cite{Pudovkin_1997, GuanLe_2016} as well as global-MHD simulations \cite{Wiltberger_2003}. As suggested by \citeA{Maltsev_1975}, \citeA{Wiltberger_2003} and \citeA{Merkin_2005}, this erosion of the magnetopause surface can be largely attributed to a proper development of Region-1 Birkeland (R1) currents and the cross tail current system which is known to reduce the dayside magnetic field strength thereby letting the solar wind penetrate further into the magnetosphere.  

 \citeA{Maltsev_1975} also suggest that such an erosion of the magnetopause is also accompanied by the equatorward motion of the polar cusps. This is also clearly seen in our comparison featured in panel (c) of figure \ref{fig:with_without_mi}. It is also additionally evident from panels (a) and (b) that the highly kinked field lines, which are a constituent of the polar cusps, are significantly more equatorward in panel (b) as compared to panel (a). The magnitude of the current density measured directly at the subsolar points for both the setups yield a value of 44 $nA m^{-2}$ for the C-0 setup whereas for the C-1 run, the value is only marginally higher at 47 $nA m^{-2}$. The bow shock is also seen to slightly move inwards for the C-1 setup, however, the most prominent difference in the shock structure appears near the polar cusps where it is seen that the bow shock is kinked significantly in the C-0 setup. This is primarily due to a stronger deflection of plasma moving in the $\pm Z$ directions by the obstacle posed by end of the polar cusp on the magnetosheath side. In panel (c), we observe that the magnetotail displays noticeable structural complexity below 10 $R_E$, but appears more diffused beyond this threshold when the MI coupling is included in the model. Furthermore, it is evident that the Kelvin-Helmholtz vortices at the magnetopause flanks differ in the two configurations. The magnetopause surface in the two configurations experiences a slightly different ambient media, therefore this is to be anticipated given the non-linear nature of the Kelvin-Helmholtz instability.

\section{FTE-Ionosphere Interaction}\label{sec:fte_iono_interaction}
 In the following sections, we carry forward our analysis with the C-1 setup that has the magnetosphere-ionosphere coupling enabled in order to study the effects of FTEs on the ionospheric surface within \texttt{MagPIE}.
\subsection{FTEs in \texttt{MagPIE} and Cusp-FTE Reconnection}
\begin{figure*}
    \centering
    \includegraphics[width=1.1\textwidth]{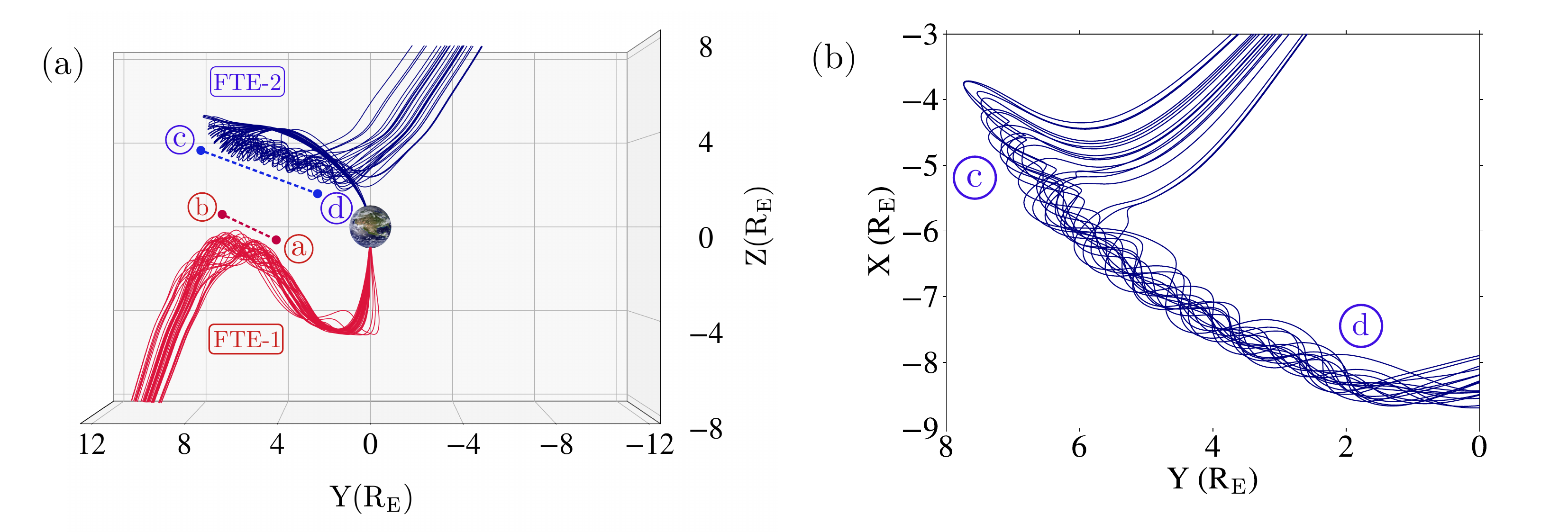}
    \caption{Panel (a) shows the magnetic field configuration using field lines traced for FTE-1 and FTE-2 at $t\sim 3225s$ and $t\sim 4905s$ respectively. The dashed lines show the approximate orientation of the flux rope axis. Panel (b) shows a top down projection of the FTE-2 field lines as they would appear looking from the zenith to the northern hemisphere.}
    \label{fig:FTE_fieldlines}
\end{figure*}

 Due to the presence of a southward IMF, the dayside magnetopause in \texttt{MagPIE} exhibits the formation of numerous FTEs. The FTEs appear as helical flux ropes, displaying the characteristic features of patchy, multiple X-line reconnection at the dayside magnetopause. These manifestations are similar to the FTEs observed in our previous resistive global-MHD simulation using the \texttt{PLUTO} code \cite{Paul_2022}.  In the following sections, we focus our attention on two such FTEs, namely FTE-1 and FTE-2. A representation of the helical magnetic field lines in the form of flux ropes corresponding to these two FTEs are shown in panel (a) of figure \ref{fig:FTE_fieldlines}. The shape and size of these FTEs evolve with time. During the epoch corresponding to panel (a) in figure \ref{fig:FTE_fieldlines}, the cross-sectional dimensions of the two FTEs were approximated as ellipses. For FTE-1, the major axis was observed to be $\sim$1.5$R_E$, while the minor axis was $\sim$0.61$R_E$. On the other hand, FTE-2 exhibited a major axis length of $\sim$2.1$R_E$ and a minor axis length of $\sim$1.55$R_E$. Additionally, the meridional (azimuthal) width of these FTEs was determined to be $\sim$56$^\circ$ for FTE-1 and $\sim$31$^\circ$ for FTE-2. The FTEs were so chosen that they lie on two different hemispheres at the pre-noon sector. Magnetic field lines due to FTE-1 connect to the southern hemisphere whereas those of FTE-2 connect to the northern hemisphere. Panel (b) in figure \ref{fig:FTE_fieldlines} shows a zoomed-in top down projection of the field lines constituting FTE-2 highlighting its azimuthal extent.
 
 The advection of these FTEs forming at the dayside magnetopause surface is governed by a superposition of magnetic tension forces and local plasma convection. Upon their motion, these FTEs generally interact with either of the polar cusps. It has been established that the FTEs impinging on the polar cusp can cause localised reconnection between the FTE flux-rope field lines and the cusp field lines owing to the favourable local topology of the magnetic field \cite{Omidi_2007}. FACs are generated within FTEs as they evolve \cite{Saunders_1984,Crooker_1990, Paul_2022}. Upon cusp-FTE reconnection, these FACs can be chanelled along the newly reconnected field lines deeper into the cusp. Such localized reconnection mechanisms have long been percieved to be prevalent and accountable for the injection of plasma into the cusps \cite{Omidi_2007}. Alternatively, the process of cusp-FTE reconnection itself may generate FACs along open as well as closed field lines. Interestingly, FACs along closed magnetospheric field lines are driven by the requirement to bring fresh magnetic field lines to the reconnection region at the cusp \cite{Kan_1996}. Both of the above processes (channelling and generation) are expected to have signatures on the ionospheric layer. Nonetheless, isolating these signatures based on their mechanism of generation remains exceedingly challenging. In the following sections, we therefore do not differentiate between the generation mechanisms, but instead, consider all possibilities and rather focus on the impact of these FTEs on the ionospheric surface.

\begin{figure}
    \centering
    \includegraphics[width=0.6\textwidth]{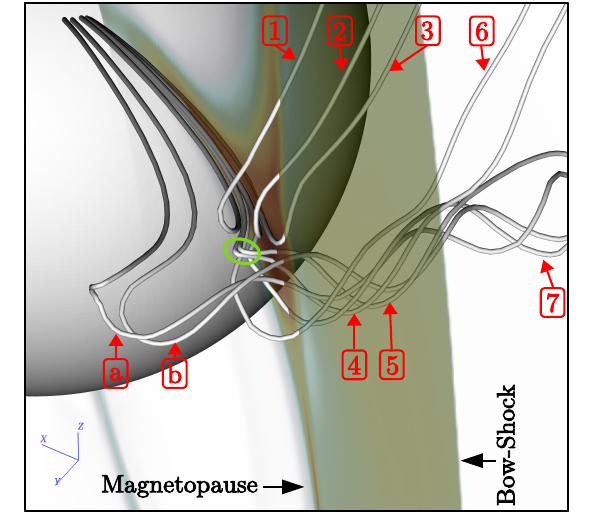}
    \caption{A zoomed in demonstration of FTE-cusp reconnection showing magnetic field lines at t$\sim$4964s. Field lines marked `a' and `b' are constituent of the FTE when it was formed earlier. Lines 1,2 and 3 are the ones originally a part of the cusp, but about to reconnect with FTE field lines, lines 4 and 5 have just reconnected with the FTE and lines 6 and 7 have reconnected with the FTE at an earlier time. The white sphere represents the internal boundary of the MHD domain. A translucent pressure pseudocolor is used to highlight the magnetopause and bow shock for reference purposes.}
    \label{fig:fte_cusp_reconect}
\end{figure}

 We first demonstrate the cusp-FTE interaction within \texttt{MagPIE} aided by several magnetic field lines corresponding to FTE-2 that are at different stages of reconnection. Figure \ref{fig:fte_cusp_reconect} shows a close-up of the magnetic field lines near the northern cusp at t$\sim$4964s. The lines labeled `a' and `b' were originally part of the FTE flux rope when it was generated at the magnetopause surface. When the FTE interacts with the cusp, a separate set of magnetic field lines that were once part of the polar cusp become part of the FTE in the following manner. The lines numbered as 1, 2 and 3 are the cusp field lines just before reconnecting with the flux rope. One can see that these lines are kinked as they approach the reconnection region. Lines 4 and 5 represent the newly reconnected field lines. They exhibit kinks similar to lines 1 and 2 as can be seen in the region enclosed by the green ellipse. However, the magnetic field lines 4 and 5 now connect through the FTE instead of connecting directly to the solar wind. Lines 6 and 7 denote the cusp field lines that have reconnected earlier than lines 4 and 5.

\begin{figure}
    \centering
    \includegraphics[width=1.0\textwidth]{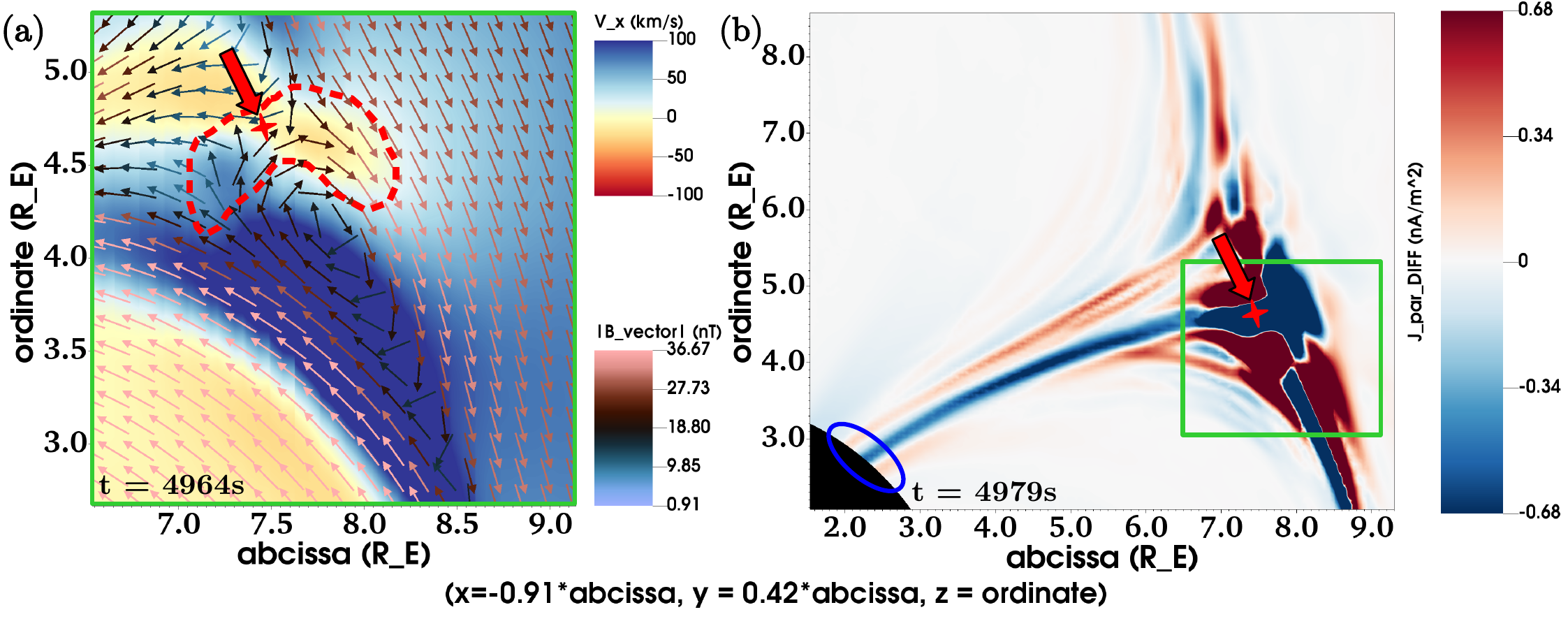}
    \caption{Panel (a) shows a zoomed in view of an oblique 2D slice exhibiting cusp-FTE reconnection at $t\sim 4964s$. The background pseudocolor represents $v_x$ whereas the superimposed thin arrows denote the local magnetic field vectors. The red `$\times$' mark highlighted by the bold red arrow denotes the location of the reconnection X-point on the 2D slices and the red dashed contour highlights the diverging reconnection exhausts from the X-point. Panel (b) represents a zoomed out view of the same region. The background pseudocolor represents the difference image of the FACs, i.e.,  $J_{\parallel DIFF}$ at $t\sim 4979s$. Once again, the red `$\times$' mark highlighted by the red arrow shows the location of the X-line. The green box in panel (b) denotes the region highlighted in panel (a). The text legend on the bottom gives the (x, y, z) coordinates based on the value of the abcissa and the ordinate values.}
    \label{fig:fte_cusp_reconect_slice}
\end{figure}

We further elaborate this cusp-FTE reconnection process through 2D slices, illustrated in Figure \ref{fig:fte_cusp_reconect_slice}. Panel (a) of figure \ref{fig:fte_cusp_reconect_slice} offers a zoomed-in view of the ongoing cusp-FTE reconnection event for FTE-2 at $t\sim 4964s$. The slice is positioned in a plane that runs parallel to the Z-axis, while being oriented at a 25$^\circ$ angle towards the positive Y-axis on the dayside. The thin arrows within panel (a) indicate the orientation of the local magnetic field vectors, revealing a field reversal region favourable for the occurence of magnetic reconnection at the position marked by a red `$\times$' symbol, highlighted by the bold red arrow. This red `$\times$' signifies the estimated location of the X-point within this specific slice. Given that the magnetic field undergoing reconnection near this X-point predominantly exhibits a $B_x$ component, we have plotted the X-component of the bulk velocities in the background pseudocolor to identify any potential reconnection exhausts. As anticipated, discernible diverging reconnection exhausts emanate from both sides of the X-point, signifying ongoing reconnection activity in that region. For the sake of clarity, we have delineated the approximate extent of these reconnection exhausts using a red dashed contour in Panel (a) of Figure \ref{fig:fte_cusp_reconect_slice}. We further note that as the cusp-FTE reconnection process itself is 3-dimensional in this case, the reconnection exhausts  are not limited to this particular slice and can flow out of plane as well.

To obtain a preliminary overview of the FAC signatures associated with cusp-FTE reconnection within the model (for a comprehensive examination of FAC signals, refer to section \ref{sec:ionospheric_signals_FTEs} later in the manuscript), we visualize the difference in FACs ($J_{\parallel DIFF (t)} = J_{\parallel (t)} - J_{\parallel (t-15s)}$) between time instances t = 4964s and t = 4979s in panel (b) of figure \ref{fig:fte_cusp_reconect_slice}. Employing this difference imaging analysis allows to isolate the impacts of transient FAC signatures associated with cusp-FTE reconnection within a dominant FAC background. The red `$\times$' mark highlighted by the thick red arrow once again denotes the location of the reconnection X-point within this frame. For further clarity, the green box in panel (b) denotes the zoomed in region highlighted in panel (a). As is evident from panel (b), FAC signals are tethered from the region of the cusp-FTE reconnection to the IB. The IB is denoted by the partial black circle on the bottom left of panel (b) and the blue ellipse highlights the region where these FAC signatures meet the IB. In the subsequent sections, these FACs are propagated to the ionospheric surface within the model in order to obtain the ionospheric signature of FTEs as presented in section \ref{sec:ionospheric_signals_FTEs}. In effect, while focusing on the polar cusp field lines, the process of cusp-FTE reconnection changes the connectivity of the magnetic field lines that were originally connected directly from the ionosphere to the solar wind to now go from the ionosphere via the FTE flux rope to the solar wind. We highlight here an important distinction between the process of standard magnetic reconnection leading to the formation of FTEs and cusp-FTE reconnection. Standard magnetic reconnection at the magnetopause occurs between IMF field lines and the field lines that are a part of the dayside closed flux. Cusp-FTE reconnection, on the other hand occurs between two sets of magnetic field lines that both constitute the net dayside open flux. From figure \ref{fig:fte_cusp_reconect} and panels (a) and (b) of figure \ref{fig:fte_cusp_reconect_slice}, it is therefore evident that FTE-cusp reconnection is a rather universal process and could serve as the key driver of ionospheric responses.

\subsection{Signatures of FTEs on the Ionosphere}\label{sec:ionospheric_signals_FTEs}
It is seen that significant perturbations are superimposed on the large scale R1 and R2 FACs after the cusp-FTE interaction. We now aim to analyse these signatures that are exhibited by the ionospheric surface in further detail. For the FTEs analysed, we first identify the flux rope from its $B_N$ component at the magnetopause surface which appears as a bipolar patch \cite{Paul_2022}. Thereafter, we locate the general region on the northern and southern cusp where the FTEs impinge on and trace the magnetic field lines from that region back to the ionospheric surface. We then look for any subsequent FAC perturbations near the footpoints of these magnetic field lines. We do not directly focus on the the exact location of the footpoints of the FTE field lines. This is because the ionospheric responses are primarily driven by the process of cusp-FTE reconnection. As such, while the FAC signals do emerge in close proximity to the footpoints of the FTE field lines, they are not always exactly located at those footpoints.

\begin{figure*}
    \centering
    \includegraphics[width=1.0\textwidth]{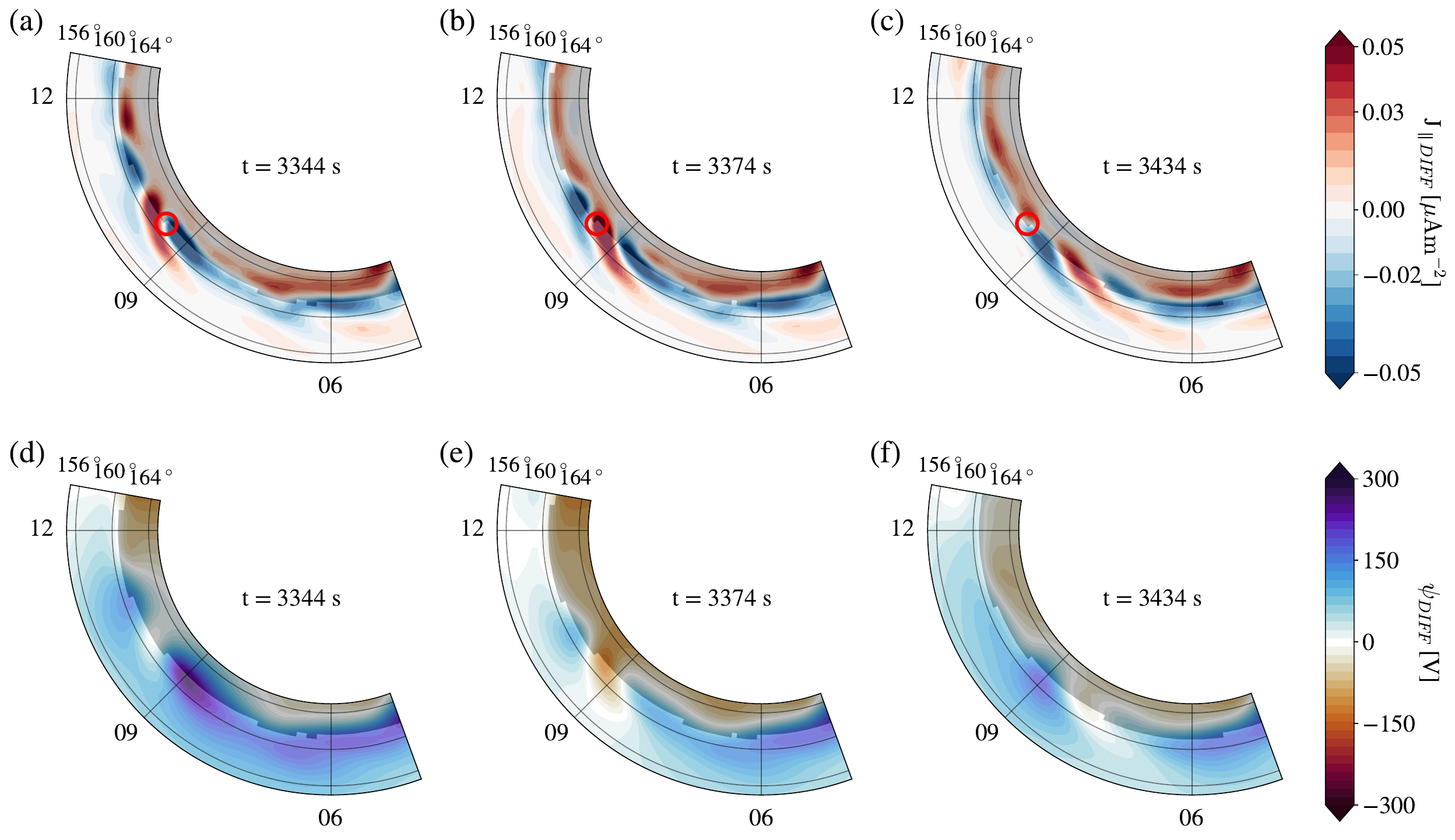}
    \caption{Background pseudocolor in panels (a) to (c) represent the difference plots $J_{\parallel DIFF}$ of the impact of FTE-1 in the ionosphere in a polar projection. Panels (d) to (f) represent the corresponding difference plots for the ionospheric potential $\psi_{DIFF}$. For all the plots in this figure, the time is denoted as inset in to the right of each panel and the light and dark shaded regions within each subplot represent the ionospheric footpoints of closed and open field lines of the magnetosphere respectively. The red circles correspond to results presented in section \ref{sec:sat_signatures}.}
    \label{fig:fte1_jpar_pot}
\end{figure*}


 The signatures associated with FTE-cusp interaction are seen to be comparatively weaker than the large scale R1 and R2 currents. In order to isolate only the effects of the FTE cusp interaction on the ionospheric surface, we employ the technique of difference image analysis. As mentioned in section \ref{sec:Initial_and_boundary}, the ionospheric FACs and the resulting potential are recalculated every 15 seconds. The difference imaging at a particular timestep is therefore given as $J_{\parallel DIFF (t)} = J_{\parallel (t)} - J_{\parallel (t-15s)}$ for the FACs and $\psi_{DIFF (t)} = \psi_{(t)} - \psi_{(t-15s)}$ for the ionospheric potential. Upon removal of the strong background current systems, the residual signal obtained from the difference imaging is taken to be explicitly due to the effect caused by an FTE on the ionospheric surface at that timestep.

Panels (a) to (c) of figure \ref{fig:fte1_jpar_pot} show the time evolution of the $J_{\parallel DIFF}$ corresponding to FTE-1 over a period of 90 seconds (we only show a small temporal subset of the evolution, the overall duration of the signal is approximately 255s). For reference, all the panels presented in figure \ref{fig:fte1_jpar_pot} correspond to the red hatched region in panel (a) of figure \ref{fig:fte_location_shade} shown in the appendix. We reiterate here that FTE-1 impinges on the pre-noon sector cusp of the southern hemisphere. We further note here that the sign of the FACs is determined with respect to the direction of magnetic field lines. As such, for the southern hemisphere, positive FACs correspond to current flowing out of the ionosphere and vice versa. The $\theta$ values correspond to geomagnetic colatitudes that start from $\theta = 0^\circ$ at the north pole and attain $\theta = 180^\circ$ at the south pole.  The $\phi$ values are such that $\phi = 0^\circ$ at  0:00 magnetic local time (midnight) and increases towards dawn having $\phi = 180^\circ$ at the 12:00 magnetic local time (noon). The $\phi$ values for the corresponding magnetic local times (MLTs) are also given in square brackets in panel (a) and (b) of figure \ref{fig:fte_location_shade} for reference and clarity.

As is evident from panel (a) of figure \ref{fig:fte1_jpar_pot}, the signal corresponding to FTE-1 appears as a tripolar patch upon first inspection. The temporal nature of the signal strength is such that it gradually increases, e.g, in panels (a) and (b), attains a temporal maxima, and then declines as shown in panel (c). We leverage this feature to determine the characteristic of this signal that travels from the cusp to the ionosphere. We take two spherical surfaces that intersects the polar cusps at two different heights, namely, 4$R_E$ and 6$R_E$ from the origin. Upon comparison of the time series of the signal corresponding to the FTE in these two cross sections, we find that the temporal maxima of the signal reaches 4$R_E$ 15 seconds later than its arrival time at 6$R_E$. Thereafter, starting at the coordinates where the maxima of the signal occurs at 6$R_E$, we trace a magnetic field line down to a height of 4$R_E$. Along each segment constituting this magnetic field line, the propagation delay `dt' of an Alfv\'en wave is calculated which depends on the local magnetic field strength and number density. The total delay ($\int_{6R_E}^{4R_E} dt$) of an Alfv\'enic signal traveling along this magnetic field line is calculated to be 14.6 seconds. We therefore conclude from this exercise that the signal corresponding to an FTE impinging at the polar cusps is communicated to the ionosphere by a signal travelling with the Alfv\'en speed along the polar cusp magnetic field lines.

Upon closer examination of the tripolar signal patches of FTE-1 throughout its evolution, as well as other FTEs displaying a comparable streak-like ionospheric FAC signature, it becomes evident that the two FAC streaks surrounding the central streak often connect on the poleward side of the FAC patch. This implies that the overall morphology of the signal essentially takes the form of an `I' shaped upward (red) FAC surrounded by a `U' shaped downward FAC pattern (blue) where the `U' shaped part can sometimes separate into two, therefore making the signal appear as a tripolar patch as shown in figure \ref{fig:fte1_jpar_pot}. We therefore name this configuration `I$\odot$U$\otimes$' where the symbols $\odot$ and $\otimes$ represent currents flowing out of and into the ionospheric surface respectively.

The corresponding $\psi_{DIFF}$ profiles are shown in panels (d) to (f). It is seen that a three lobe perturbation arises in the potential with a negative lobe surrounded by positive potential lobes on both sides. This three lobe configuration gradually appears in panel (d), rising in strength, and eventually disintegrating after panel (f). In the panels (a) to (f), slightly darker shades towards the poleward half of each plot represents the region of open field lines and the remaining portion denotes the closed field line region. As such, the interface between the dark and lightly shaded regions represent the open-closed magnetic field line boundary (OCB) of the magnetosphere at those timesteps. We find that the spatial maxima of the signal travels in the $\phi$ direction close to the OCB.

\begin{figure*}
    \centering
    \includegraphics[width=1.0\textwidth]{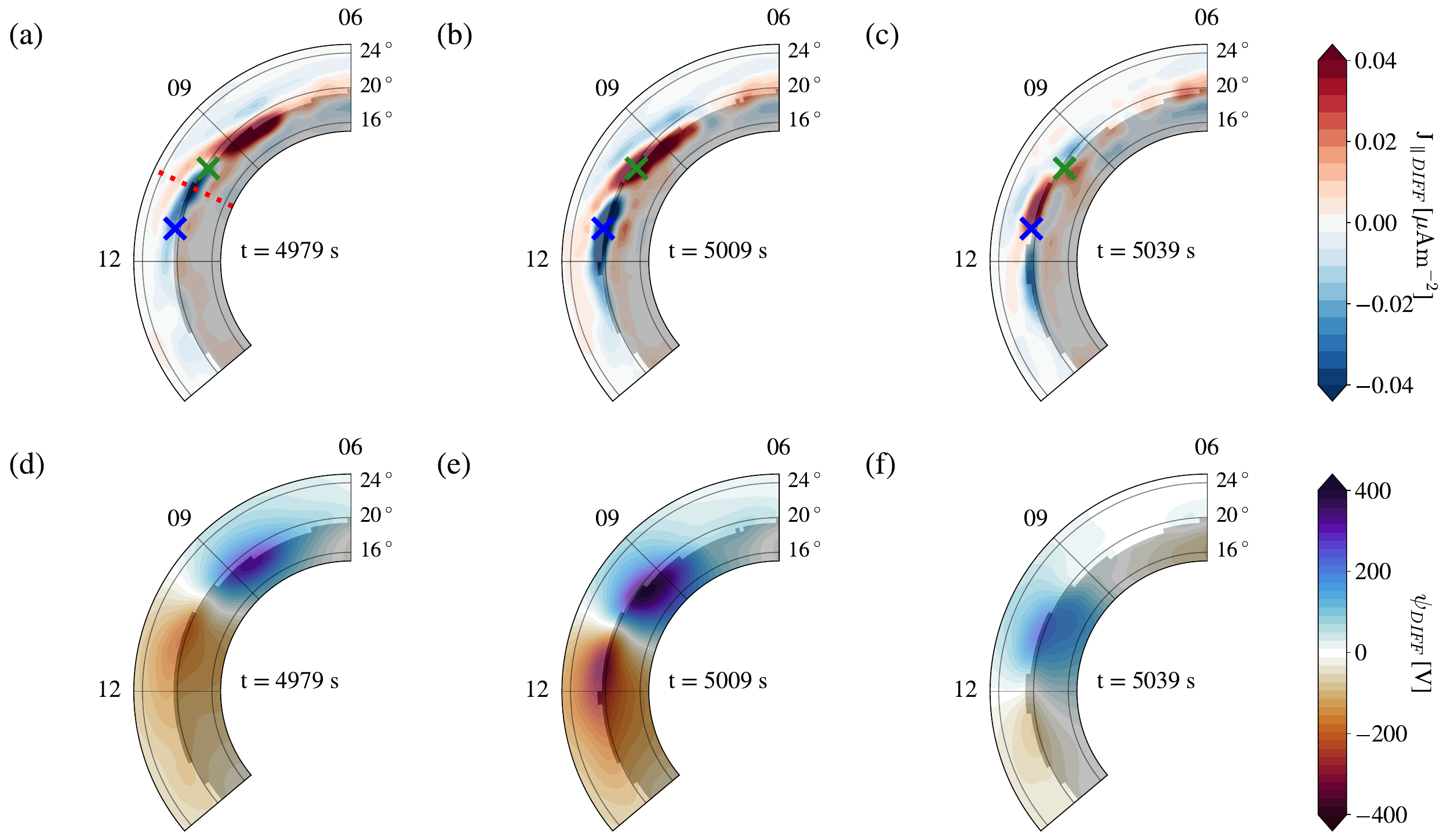}
    \caption{Same as figure \ref{fig:fte1_jpar_pot} but for FTE-2. The red dotted line in panel (a) represents the slices presented in figure \ref{fig:fte_cusp_reconect_slice}. The blue and green `$\times$' marks correspond to results presented in section \ref{sec:sat_signatures}.}
    \label{fig:fte2_jpar_pot}
\end{figure*}

Panels (a) to (c) of figure \ref{fig:fte2_jpar_pot} similarly show the time evolution of the $J_{\parallel DIFF}$ corresponding to FTE-2 over a period of 60 seconds. The region featured in all the subplots of figure \ref{fig:fte2_jpar_pot} corresponds to the blue hatched region in panel (b) of figure \ref{fig:fte_location_shade} from the appendix. FTE-2 also impinges on the pre-noon sector but at the cusp of the northern hemisphere. As the FAC signs are defined based on the direction of the magnetic field, care must be taken for the northern hemisphere as the currents flowing out of the ionosphere have a negative sign, whereas currents flowing into the ionosphere are positive. Evident from panels (a) to (c), the morphology of the signal generated by FTE-2 is clearly a combination of an `I' shaped patch surrounded by a `U' shaped patch. In the case of FTE-2, the `I' shaped current is the one that is flowing out of the ionosphere, with the `U' shaped portion flowing into the ionosphere. It is therefore apparent that this configuration of `I$\odot$U$\otimes$' in the FACs due to FTE-2 are exactly same as what was observed for the case of FTE-1. Due to the fact that FACs of the same sign can have different meanings depending on the northern and southern ionospheric surface, the sense of directionality of the the FACs have been summarised in table \ref{tab:FTE_FAC_sense}.  

\begin{table}[]
\begin{center}
\begin{tabular}{ccccc}
\hline\hline
      & \multicolumn{1}{c}{Pre-Noon} & \multicolumn{1}{c}{Post-Noon}  \\ \hline
NH & I$\odot$U$\otimes$[R1$\otimes$R2$\odot$]&I$\otimes$U$\odot$[R1$\odot$R2$\otimes$] \\
SH & I$\odot$U$\otimes$[R1$\otimes$R2$\odot$]&I$\otimes$U$\odot$[R1$\odot$R2$\otimes$] \\\hline\hline
\end{tabular}%
\caption{Direction of field aligned currents}
\label{tab:FTE_FAC_sense}
\end{center}
\end{table}

The red dashed line in panel (a) of figure \ref{fig:fte2_jpar_pot} intersects the FAC signal related to FTE-2 radially. This red dashed line actually corresponds to the meridional slice presented in panel (b) of figure \ref{fig:fte_cusp_reconect_slice}. This serves to further substantiate that the FAC signal shown in panel (b) of figure \ref{fig:fte_cusp_reconect_slice}, that connects from the polar cusp to the IB is what essentially constitutes the morphology of the `I' and `U' shaped patches seen on the ionospheric surface in figure \ref{fig:fte2_jpar_pot}. One can easily relate the blue structure surrounded on both sides by the red structures in the region enclosed by the blue ellipse in panel (b) of figure \ref{fig:fte_cusp_reconect_slice} to be a part of the blue patch surrounded by the red patch in panel (a) of figure \ref{fig:fte2_jpar_pot}.

It is seen that for the IMF considered in this study, the pre-noon sector of both, the northern and southern hemispheres has a I$\odot$U$\otimes$ configuration, whereas the post-noon sector has a I$\otimes$U$\odot$ sense. Thus, given a particular hemisphere, the sense of the `I' and `U' shaped FACs are reversed in the pre-noon sector with respect to the post-noon sector. Table \ref{tab:FTE_FAC_sense} also summarises in square braces, the sense of the large scale R1 and R2 currents in the respective sectors. The FTEs at the magnetopause surface are related to the R1 current system. One can therefore draw an interesting correlation that the `I' portion of the FTE signature is always in the opposite sense of the local R1 current system in both hemispheres, whereas, the `U' portion has a similar sense. This leads to the inference that the `I' portion of the FTE signal depletes the R1 current system whereas the `U' portion enhances it. The potential pattern generated by FTE-2 shows a two lobe structure instead of the three-lobe structure seen for FTE-1. This two lobe potential pattern also travels in the $\phi$ direction with the movement of the $J_{\parallel DIFF}$ signature as seen from panels (d) to (f) of figure \ref{fig:fte2_jpar_pot}. 


It is interesting to note that the combined `U' and `I' shaped patches of both the FTEs exhibit an azimuthal motion. This indicates that the signal arises from different regions on the magnetosheath side of the polar cusps at different times. We therefore reinforce our argument here as to the reason why these typical signals are specifically due to the cusp-FTE interaction. First of all, as the signal itself is seen to propagate, we only look at possible causes that are transient, e.g, FTE motion through the magnetopause surface, X-line spreading, and cusp-FTE interaction. As soon as the signal patch appears on the ionospheric surface, a local spatial maximum exists within this signal patch. The existence of such a maxima that propagates along with the signal patch is unlikely to occur if the signal were solely due to the perturbation caused by the motion of the FTE along the magnetopause surface. On the contrary, the most general situation regarding X-line spreading at the magnetopause suggests a bidirectional spread. This implies that once an X-line begins to form, it extends along both the dawnward and duskward directions simultaneously. Consequently, if the ionospheric signal were solely attributed to this X-line expansion, the ionospheric patch would initially appear at a specific location and subsequently spread both towards dawn and dusk concurrently. However, this is not what is seen from our model. We find that the direction of propagation of the FTE signals for all the FTEs generated within the model is strictly related to the spatial orientation of the flux rope on the magnetopause surface and the direction of propagation of the signal is solely determined by which sections of the FTE progressively interacts with the corresponding polar cusps.
 
 The red dashed line in panel (a) of figure \ref{fig:FTE_fieldlines} denotes the approximate orientation of the axis of the flux rope corresponding to FTE-1. This orientation has also been verified by checking the bipolar $B_N$ component on the magnetopause surface for this FTE. It is seen that the axis of FTE-1 is tilted slightly. As the FTE plunges onto the southern cusp, the end marked as `a' will interact with the cusp earlier than the end marked `b'. Therefore, the signal due to this impact will start at a region of higher $\phi$ value (higher MLT/noon side) and gradually travel towards lower $\phi$ values (lower MLT/dawn side) as the rest of the FTE flux rope interacts up to the end marked  as `b'. This is exactly what is seen from figure \ref{fig:fte1_jpar_pot}. Similarly, for FTE-2, the side marked as `c' in panel (a) of figure \ref{fig:FTE_fieldlines} will interact with the cusp earlier. The points `c' and `d' of panel (a) for FTE-2 are also marked in panel (b) of figure \ref{fig:FTE_fieldlines} that shows a top-down zoomed in view of FTE-2. It is evident from the azimuthal separation of the points `c' and `d' that upon impinging on the cusp, the signal due to cusp-FTE interaction will travel from `c' towards `d'.Therefore the signal is initiated at a lower value of $\phi$ (lower MLT) and progressively propagates towards higher $\phi$ values (higher MLT) as can be seen in figure \ref{fig:fte2_jpar_pot}.

\subsection{Auroral Signatures and Ionospheric Vortices} \label{sec:auroral_sig}
It has been proposed that any feature outlined at the magnetopause surface will have a mapping distortion associated with it when mapped along magnetic field lines down to ionospheric heights. \citeA{Crooker_1990} highlighted that the pattern of flux tube footpoints, when projected on the ionosphere, resemble the discrete auroral arcs that are generally seen in the mid-day auroral oval. \citeA{Sandholt_1986} proposed that such arcs maybe the result of a local injection of magnetosheath plasma at the polar cusps mediated by FTEs. A special feature of these discrete auroral arcs are the fact that they appear as fan-shaped structures that are radially striated and focusing towards the cusp (see figure 5 from \citeA{Crooker_1990} or figure 6 from \citeA{Lundin_1985}). Panels (a) to (c) of figure \ref{fig:fte1_jpar_pot} shows the $J_{\parallel DIFF}$ produced by FTE-1 throughout its evolution. Upward FACs are generally associated with the formation of auroral arcs. Panels (a) to (c) of figure \ref{fig:fte1_jpar_pot} represents the southern hemisphere and therefore, upwards FACs have a positive value. We see from figure \ref{fig:fte1_jpar_pot} that a very fine resemblance exists between the FTE-1 signature on the ionospheric surface with the aforementioned auroral arcs presented by \citeA{Crooker_1990}. Panels (a) to (c) of  figure \ref{fig:fte1_jpar_pot} focusses on the FTE signatures at the pre-noon sector of the southern hemisphere. Due to the symmetry in our simulation, FTE signatures in the pre-noon sector of the southern hemisphere are similar to the post-noon sector of the northern hemisphere and vice-versa. As such, such fan shaped arcs are also found to be abundant in the post noon-sector of the northern hemisphere. Indeed \citeA{Crooker_1990} highlights the abundance of afternoon arcs in the northern hemisphere for a negative $B_Y$ component of the IMF similar to our present study. The FACs created by cusp-FTE interaction can therefore be attributed to be one of the factors leading to their generation. 

\begin{figure}
    \centering
    \includegraphics[width=0.4\textwidth]{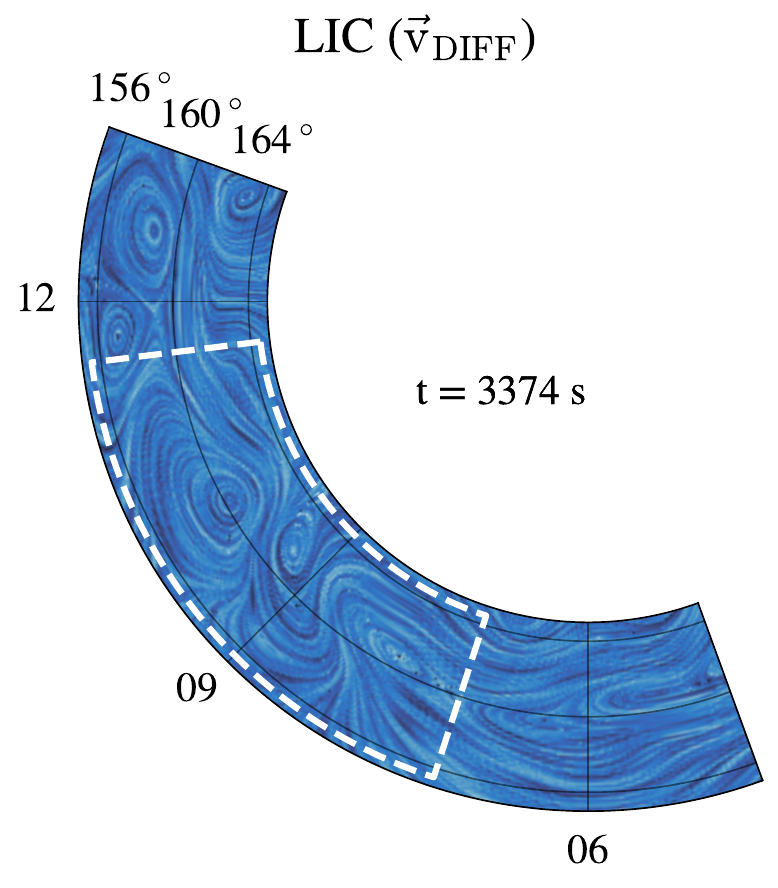}
    \caption{A plot of the line integral convolution of the velocity field on the ionospheric surface produced by the FACs due to FTE-2 at t $\sim$ 5009s. The region surrounded by the white dashed lines highlight the generated vortex pattern. }
    \label{fig:fte_TCV}
\end{figure}

Panels (a) to (c) of figure \ref{fig:fte2_jpar_pot} is the $J_{\parallel DIFF}$ produced by FTE-2  throughout its evolution. FTE-2 interacts with the northern cusp and therefore, panels (a) to (c) are a representation of the northern ionosphere. The upwards currents, in this case, are negative in sign. As seen in these panels, FTEs in the pre-noon sector of the northern hemisphere (or post-noon sector in the southern hemisphere) could also lead to arc formation in a similar manner. However, a comparison of the structures in figure \ref{fig:fte1_jpar_pot} and figure \ref{fig:fte2_jpar_pot} clearly highlights the dissimilarity in their spatial orientations. The discrete arcs due to FTE-1 would appear near the OCB and fan out well into the closed field line region. For FTE-2, however, the arc would appear primarily along the OCB. Adapting from \citeA{Crooker_1990}, with the help of magnetic field line tracing, we have verified that the morphology of the FACs due to FTEs as seen in panels (a) to (c) of figures \ref{fig:fte1_jpar_pot} and \ref{fig:fte2_jpar_pot} are purely due to mapping distortions caused by the large scale magnetospheric topology of open and closed field lines near the magnetopause boundary layer.

We contend that despite figures \ref{fig:fte1_jpar_pot} and \ref{fig:fte2_jpar_pot} depicting difference images, they can serve as a viable representation of localized auroral intensifications. As mentioned before, upwards FACs are generally associated with arc formation and upwards FACs in the southern hemisphere would be positive in sign. Considering the FAC signatures of FTE-1 in figure \ref{fig:fte1_jpar_pot}, a positive patch in the difference image could mean one of two things: (a) a localised decrease in the negative (downward) FAC content in that region or (b) a localised increase in the positive(upward) FAC content in that region. Scenario (a) is equivalent to the superimposition of the pre existing large-scale downward FACs with a localised upwards FAC patch due to cusp-FTE reconnection. Scenario (b) would directly lead to a localised increase in the upwards FACs. Thus, both scenarios could be ascribed to localized intensification of auroral emissions, thereby generating arcs. A comparable perspective can be extended to FTE-2 as well.

Travelling convection vortices (TCVs) are typically observed at the high latitude ionosphere. Among its several generation mechanisms, their generation by flux transfer events is one of the widely accepted notions. Panel (c) of figure \ref{fig:fte_TCV} shows the line integral convolution of the velocity field generated by the FACs shown in panel (b) of figure \ref{fig:fte1_jpar_pot} corresponding to FTE-1 at t=3374s. The velocity pattern generated solely due to the FTE FACs has been isolated as $\vec{v}_{DIFF} = (-\nabla\psi_{DIFF}\times\mathbf{B})/B^2$. As can be clearly seen in the region bounded by the white dashed line, the FACs give rise to three distinct vortices between 7 MLT to 11 MLT. The three vortex centers lie at 160.5$^\circ$, 161$^\circ$ and 160$^\circ$ respectively while moving clockwise in MLT. Consistent with observations,  this translates to latitudes of 70.5$^\circ$, 71$^\circ$ and 70$^\circ$ respectively \cite{Amm_2002}. For the case of FTE-2, a twin vortex pattern is seen (plot not shown). The vortices have an alternating direction where upward FACs are associated with clockwise vortices and downward FACs are associated with anti-clockwise vortices. As the FACs associated with the formation of these vortices are transient, these vortices act as perturbations and are expected to be superimposed and advected with the large scale plasma flows of the ionosphere.

\section{Detectable Satellite Signatures}\label{sec:sat_signatures}

\begin{table}[]
\begin{center}
\begin{tabular}{llll}\hline\hline
                       & Sat-1 & Sat-2.1 & Sat-2.2 \\\hline
$\theta$ {[}degrees{]} & 161.3 & 19.6    & 20.6    \\
$\phi$ {[}degrees{]}   & 142.7 & 146.9   & 169.4   \\\hline\hline
\end{tabular}
\caption{Location of Virtual Satellites}
\label{tab:sat_location}
\end{center}
\end{table}

In this particular section, our focus is on providing a detailed analysis of how the transient characteristics of the FACs generated by these FTEs would be observed by in-situ probes that are positioned suitably at the right place and time. More specifically, we look into time series spacecraft signatures that would be detected by ionospheric satellites with a high time cadence e.g, SWARM which is a set of three satellites (SWARM-A, SWARM-B and SWARM-C) that traverse the ionosphere and thermosphere at altitudes between 450 to 530 km 
 in polar orbits \cite{Ritter2013}. 

The specific coordinates of the individual virtual satellites for each of the FTEs have been presented in table \ref{tab:sat_location}. The probe named `Sat-1' corresponds to measurements of FTE-1 whereas the probes named `Sat-2.1' and `Sat-2.2' correspond to measurements for FTE-2. The location of the satellites are also highlighted as appropriately colored circles (for FTE-1) and crosses (for FTE-2) overplotted on the FAC background in figures \ref{fig:fte1_jpar_pot} and \ref{fig:fte2_jpar_pot}. The red circles in panels (a) to (c) of figure \ref{fig:fte1_jpar_pot} represents Sat-1 whereas the green and blue `$\times$' marks in panel (a) to (c) of figure \ref{fig:fte2_jpar_pot} represent Sat-2.1 and Sat-2.2 respectively. The positioning of the satellites depicted in all three subplots also provides insights into  how the FAC patches traverse through the virtual in-situ probes.

Considering a general linear speed of the SWARM spacecrafts to be 7.1 kms$^{-1}$, at an altitude of 450 km, this yields an angular speed of 5.95$\times 10^{-2}$degrees/second. As far as the ionospheric grids are concerned, this means that the spacecraft crosses one latitudinal grid cell in approximately 9 seconds. This is indeed smaller than the total time duration showcased in figures \ref{fig:fte1_jpar_pot} and \ref{fig:fte2_jpar_pot}. However, quite a few complications arise when considering a moving virtual satellite. For polar orbits, the virtual satellite can have multiple directions of approach with respect to the signal of interest and the choice of any particular direction would be arbitrary. This would mean that due to the spatial morphology of the signal, significant variations in the observed time series may arise depending on the relative trajectory of the FAC patch and the spacecraft. Additionally, for moving virtual satellites, depending on the direction of motion, the perceived signal may be temporally distorted. Therefore, for simplicity and also as a way to generalise the temporal signature perceived by an in-situ probe, we consider stationary satellites in the following analysis.     
 
\begin{figure*}
    \centering
    \includegraphics[width=1.0\textwidth]{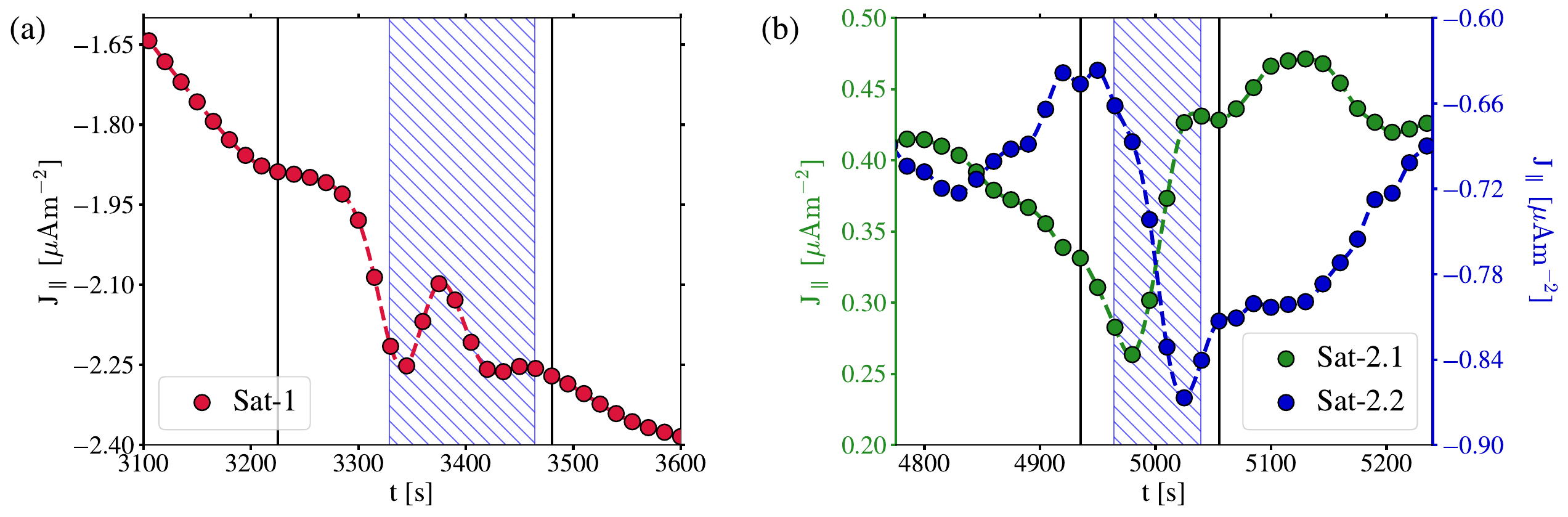}
    \caption{Panel (a) shows the FAC time series produced by FTE-1 detected by a satellite located at the red circle in panels (a) to (c) of figure \ref{fig:fte1_jpar_pot}. The green and blue curves in panel (b) represent time series of two different satellites located at the correspondingly colored `$\times$' marks in panels (a) to (c)  of figure \ref{fig:fte2_jpar_pot}. The blue shaded portions in panels (a) and (b) represent the time duration highlighted in figures \ref{fig:fte1_jpar_pot} (for FTE-1) and \ref{fig:fte2_jpar_pot} (for FTE-2) respectively.}
    \label{fig:satsign_both}
\end{figure*}

 Panel (b) of figure \ref{fig:satsign_both} shows the satellite signatures due to FTE-2 on the northern ionosphere. For FTE-2, we employ two virtual satellites at the locations given by the green and blue `$\times$' marks in panels (a) to (c) of figure \ref{fig:fte2_jpar_pot}. Note that unlike FTE-1, FTE-2 forms at a much later time when the equatorward motion of the large scale R1 and R2 currents have halted. The location of the two virtual satellites have been so chosen that one of them lies at the maximum value (Sat-2.1) obtained during the time evolution, a part of which is shown in figure \ref{fig:fte2_jpar_pot}, and the other at the minima (Sat-2.2). For FTE-2, the signal from the FTE-cusp interaction can be clearly discerned between t$\sim$4935s and t$\sim$5055s on the ionospheric surface. Similar to panel (a), this time duration is encased between the two solid black vertical lines in panel (b) of figure \ref{fig:satsign_both}. As seen by the green and blue curves which represent signals from Sat-2.1 and Sat-2.2 respectively, the signal can have vastly different morphologies depending strongly on the coordinates of observation. Sat-2.1 lies at the outer radial edge of the R-1 current arc having a positive value (sign of FACs are based on the direction of magnetic field lines). This probe initially encounters the negative portion of the FAC patch seen in the difference images of figure \ref{fig:fte2_jpar_pot}. This manifests as an initial decrease in the FAC value. Soon thereafter, the positive portion of the signal passes through Sat-2.1 and this results in a rapid rise in the FAC value.

 As previously stated, Sat-2.2 is chosen to be at the minima of the signal due to FTE-2. This sets Sat-2.2 at the edge of the negatively signed R1 FAC in the northern hemisphere. As expected due to the spatial morphology of the signal, at the location of Sat-2.2, the FAC shows a rapid decrease due to the negative portion of the signal passing through it and following this, a slight increase when the probe encounters the positive segment. Considering the time series given by Sat-2.1 (and Sat-2.2), it is difficult to pinpoint with absolute certainty the reason that the FAC magnitude starts to fall(rise) at approximately t$\sim$4850s which is much earlier than the time the FTE has first contact with the polar cusps (at t$\sim$4905s). One of the probable reasons for the decrease is that as the FTE gains speed during its motion along the magnetopause surface, it displaces plasma located at the forefront of the FTE. This perturbation may travel along the magnetic field lines and eventually manifest as a gradual decrease(increase) in the FAC magnitude at that particular location just before FTE-2 interacts with the cusp. A more probable reason could be that the change is due to a residual signal left after a severe encounter with a previous FTE that impinges on the cusp at t$\sim$ 4695s.

\begin{figure*}
    \centering
    \includegraphics[width=1.0\textwidth]{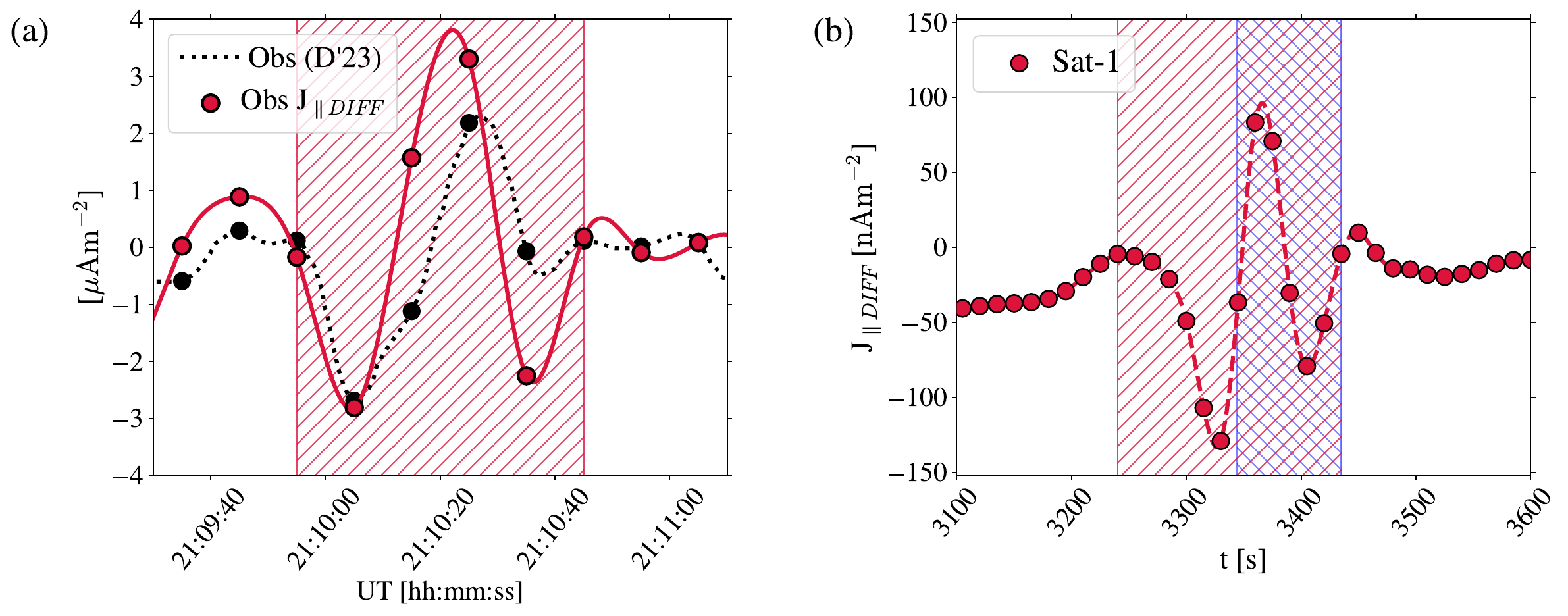}
    \caption{The black dotted line in panel (a) corresponds to an observation by \citeA{Dong_2023}. The black circles on the dotted curve represent the points where the data has been resampled at. The solid red dots in panel (a) represents the $J_{\parallel DIFF}$ obtained from the resampled observational data and the red solid line represents a spline fit to the scatter points. Panel (b) corresponds to the $J_{\parallel DIFF}$ obtained for FTE-1 at the location denoted by the red circles in panels (a) to (c) of figure \ref{fig:fte1_jpar_pot}.}
    \label{fig:obs_fte1}
\end{figure*}

 We now focus on a recent study by \citeA{Dong_2023} that reports the simultaneous observations of negative-positive FAC pairs by the CLUSTER and SWARM spacecrafts. Two FAC pairs were observed by the SWARM spacecrafts starting at approximately 21:08 UT and 21:10 UT on October 7, 2015. We add a note of caution here that the following section does not present a one to one event analysis. Instead, the goal is to highlight the morphological similarity between the observed FACs at the heights of SWARM spacecrafts and those detected as the ionospheric signature of FTEs in our simulation.

 For purposes of generalisation, we choose the second FAC pair observed by \citeA{Dong_2023} beginning at approximately 21:10 UT. This FAC pair has a typical structure with a negative peak followed by a positive peak in the FAC magnitude. The black dotted curve in panel (a) of figure \ref{fig:obs_fte1} represents the FAC pair highlighted in panel (g) of figure 2 in \citeA{Dong_2023}. In order to draw comparisons with the simulated signatures, we first resample the observed data at intervals of 10 seconds. The black circles on the dotted curve in panel (a) of figure \ref{fig:obs_fte1} represents the location where the resampled data points are located. We note here that we have resampled the data at intervals of 10 seconds instead of the 15 seconds time cadence that our simulation has. This has been done with the aim to adequately capture the fall and rise in the observed FAC pair. As evident by the location of the black dots in panel (a) of figure \ref{fig:obs_fte1}, resampling at a 10 second cadence ensures that one of the data points lie on the positive peak and one at the negative peak of the observed data with at least one data point in between these two peaks. Now, as mentioned in \citeA{Dong_2023}, the plotted FAC pair is obtained after the subtraction of the CHAOS-6 magnetic model, i.e $J_{\parallel t}$ = $J_{\parallel t (SWARM)}$ - $J_{\parallel t (CHAOS6)}$. One can therefore easily derive that $J_{\parallel t}$ - $J_{\parallel t-10s}$ = $J_{\parallel t (SWARM)}$ - $J_{\parallel t-10s (SWARM)}$. This means that a consecutive difference of the resampled FAC data from the plot is equal to the consecutive difference of the total FACs detected by SWARM. This is analogous to the consecutive difference($J_{\parallel DIFF}$) obtained in our difference image analysis. The red circles in panel (a) of figure \ref{fig:obs_fte1} represent this consecutive difference (Obs$J_{\parallel DIFF}$ = $J_{\parallel t}$ - $J_{\parallel t-10s}$) of the observational data from SWARM. The solid red line is a spline fit to the scatter points. Panel (b) of figure \ref{fig:obs_fte1} shows an in-situ measurement obtained at the location of Sat-1 if one measures the $J_{\parallel DIFF}$ instead of just $J_{\parallel}$. Upon comparison of the red hatched region in panels (a) and (b) of figure \ref{fig:obs_fte1}, the similarities in morphology between the two signal patterns are evident. The interaction of FTEs with the cusp may be one of the generation mechanisms of such an FAC pair. Even though \citeA{Dong_2023} analyse these FACS to be bipolar with a positive and a negative end, upon careful observation of the observational data, we also see a very weak negative FAC patch detected just after the positive portion. Therefore, morphologically speaking, the signal is very similar to the I$\odot$U$\otimes$ seen during the difference image analysis of FTE-1, i.e., as seen in panels (a) to (c) of figure \ref{fig:fte1_jpar_pot}. The blue hatched region in panel (b) of figure \ref{fig:obs_fte1} represents the duration highlighted in panels (a) to (c) of figure \ref{fig:fte1_jpar_pot} for reference. \citeA{Dong_2023} report the speed of the FAC pair to be 0.8km/s. Considering the height of SWARM to be 450km during that time, the angular speed is calculated to be 1.1$\times10^{-4}$ radians/second. Upon analysing the movement of the signal from the simulation between t$\sim$3345s and t$\sim$3390s, the estimated total angular speed (calculated as $\sqrt{\Delta\theta^2 + \Delta\phi^2}$/$\Delta$t) of the signal due to the FTE is calculated to be 4.9$\times10^{-3}$ radians/second which is about an order of magnitude higher. We reiterate here that the aim of the above comparison is not to extract a one-to-one event based output. Instead, it is to highlight the that such FACs at SWARM heights could in fact be produced by FTE-cusp interaction like the ones presented in this study. As such, some quantitative differences are expected.

\section{Discussions and Summary}\label{sec:disc_and_summ}

This study presents a comprehensive examination of the effects of flux transfer events on the ionospheric surface. In order to accomplish this, we have incorporated a two-way coupled ionospheric model into a previously developed magnetospheric model to form a combined magnetosphere-ionosphere (MI) model in the open source PLUTO code. The combined model has been named \texttt{MagPIE} (\textit{\textbf{Mag}netosphere of \textbf{P}lanets with \textbf{I}onospheric \textbf{E}lectrostatics}). It is seen that a significant difference in the large scale magnetospheric shape appears upon the incorporation of the two way coupled MI module that is consistent with observations during SIMF conditions. The most prominent of these differences are magnetopause erosion and equatorward motion of the polar cusps as seen in figure \ref{fig:with_without_mi}. 

Two FTEs, namely FTE-1 and FTE-2 were selected in order for a detailed analysis with respect to the signatures they administer on the ionospheric layer. One must exercise caution in interpreting the FAC signatures from the consecutive difference images presented in figures \ref{fig:fte1_jpar_pot} and \ref{fig:fte2_jpar_pot}, most importantly, the morphology of the FAC patches. The signal that is generated by an FTE manifests itself on the surface of the ionosphere as a `I' shaped patch that is surrounded by a `U' shaped patch of the opposite sign. A significant portion of the trailing patch of the total FAC morphology may be due to the contribution from the movement of the `I' shaped patch away from that region (as the entire structure is in motion). In some cases, the trailing patch may almost entirely be formed by the moving away of the `I' shaped patch. In such cases, even a positive-negative FAC pair would appear to be a three patch structure in the difference-images. As such, the `I' shaped patch may also be slightly modified by the leading edge of the `U' shaped patch. However, for the FTEs presented in this study, we find that the morphology (`I' patch surrounded by `U' patch) remains unmodified upon subtraction of the background of only the first timestep (t = 3329 for FTE-1 and t = 4964 for FTE-2) from all the frames. It is just the magnitude of the patches that change. In any case, such a consecutive difference image analysis is indeed found to be the clearest way of visualising the signatures produced due to FTEs on the ionosphere.

Consistent with the proposition of \citeA{Glassmeier_1996}, the effects due to cusp-FTE reconnection is found to propagate from the polar cusps towards the IB of the simulation with the speed of an Alfv\'en wave. The simple flow pattern envisaged by \citeA{Southwood_1985} was that of a twin vortex on the ionosphere produced in response to a pair of FACs. Our study reveals that such FAC pairs are indeed formed by FTEs, be it in a combination of `I' and `U' shaped patches as highlighted in section \ref{sec:fte_iono_interaction}. It however remains to be seen whether the FAC pairs are simply channelled through the ionosphere upon cusp-FTE reconnection or are they a result of the reconnection process itself. We highlight a very important point here that, contrary to the proposition that the FACs seen on the ionosphere are a direct consequence of reconnection at the dayside magnetopause, the results from our simulation assert that it is in fact the cusp-FTE reconnection that causes the dominant responses on the ionosphere by channelling (and possibly generation) of strong FAC patches that travel along the cusp field lines.

It is also interesting to note that the extent of the FACs due to the FTEs greatly surpass the region covered by the FTE footpoints on the ionosphere. As such, the resulting potential and eventually, the convection set up in response to these FACs are also spread out over a much broader region than the coverage of the FTE footpoints on the ionosphere. We reiterate here that the signal due to the cusp-FTE interaction in our simulation is seen to travels close to the footpoints of the magnetic field lines of the FTEs on the ionospheric surface. This is, however, not a strict requirement. As the channelling mechanism of the FACs is the process of cusp-FTE reconnection, the signal is expected to be located at the footpoints of the cusp field lines that reconnect with the flux rope instead of the footpoints of the flux rope field lines themselves. A few studies have reported the simultaneous observation of FTEs in the high altitude cusp and their corresponding responses in the polar ionosphere \cite{Marchaudon_2004,Dong_2023}.

Also consistent with \citeA{Southwood_1985}, the effect of an FTE does manifest itself as vortex-like ionospheric flow patterns. The vortex pattern shown in figure \ref{fig:fte_TCV} is expected to superimposed on the large scale flow. This is consistent with observations by \citeA{Oskavik_2004} that suggest that the flow pattern due to FTEs would appear to be a ripple into the large scale ionospheric convection pattern. However, the flow is not necessarily a twin-vortex. We find that depending on the morphology of the FACs, the flow can either be a twin-vortex or a combination of more than two vortices as seen from figure \ref{fig:fte_TCV}. Either way, there would exist enhanced flow channels between the vortices as seen in the cusp ionosphere \cite{Pinnock_1993,Rodger_1997,Neudegg_2000}. Such vortices are named as travelling convection vortices(TCVs) in literature and despite the fact that a number of generation mechanisms have been proposed as the cause of these TCVs, including FTEs, no definitive explanation had yet been provided \cite{Moretto_1998,Kim_2017}. Our simulations explicitly show how FTEs can actually generate such vortices. 

As analysed in section \ref{sec:fte_iono_interaction} the propagation of the signal is seen to be directly correlated to the orientation of the FTE flux rope on the magnetopause surface. Figure \ref{fig:fte_TCV} demonstrates that FTEs can indeed be responsible for the generation of multiple vortex pairs on the ionospheric surface, which may then advect with the large scale ionospheric flow pattern. An enhanced flow patch exists in the region between any two vortices. Given that the direction of propagation of the signal on the ionospheric surface is related to the global orientation of the FTE, one can, in principle, infer the past orientation of the FTE on the magnetopause surface based on the longitudinal propagation of the enhanced flow channel as derived from ground based radars such as SuperDARN.

The polar projection of the FACs related to these FTEs are found to have an interesting resemblance with discrete dayside auroral arcs commonly observed to be embedded within the diffuse aurorae \cite{Crooker_1990}. \citeA{Lundin_1985} and \citeA{Crooker_1990} proposed that such arcs may be produced by a local injection of magnetosheath plasma inside the dayside boundary layer by the process of magnetic reconnection. Our work relates well with this proposition, and it is seen that the process of FTE-ionosphere interaction can indeed be a probable cause for the formation of these discrete auroral arcs. It has also been highlighted by \citeA{Crooker_1990} that the discrete arcs are prevalent in the early afternoon sector of the northern hemisphere for a negative IMF $B_Y$ component. This is consistent with the inferences from our simulation that the discrete arcs lie mainly in the afternoon sector of the northern hemisphere. Due to the symmetry in the magnetic field lines of the system, the arcs in the southern hemisphere occur in the pre-noon sector as seen in panels (a) to (c) of figure \ref{fig:fte1_jpar_pot}. However, we add for completeness that, consistent with \citeA{Lundin_1985}, the pre-noon sector of the northern hemisphere (and the post-noon sector of the southern hemisphere) also, very rarely, show radially striated arcs similar to panel (a), be it of a very limited local time extent. The region of appearance of these discrete arcs would be reversed (pre-noon in the NH and post-noon in the SH) upon reversal of the IMF $B_Y$. In any case, FTEs appear responsible for the generation of discrete radially striated dayside auroral arcs as envisaged by \citeA{Crooker_1990}.

We also employ virtual satellites in the ionospheric domain in order to infer time series signatures of the FACs produced due to the FTEs. Considering stationary satellites, the FAC signatures are found to be transient and enclosed within a duration of approximately 255 seconds for FTE-1 and 120 seconds for FTE-2 in the virtual in-situ data. The SWARM constellation has been vital towards mapping the near Earth FACs in a high temporal and spatial resolution \cite{Ritter2013}. Given the accuracy specified in the SWARM vector magnetic field data at 0.15nT and an average spacecraft separation of 40km, the achievable resolution in FAC density is calculated to be $\pm$0.004$\mu Am^{-2}$. In principle, this is more than sufficient to measure the rapid changes of 0.2 to 0.3 $\mu Am^{-2}$ in FACs. A spatial resolution of 1km can be obtained by the full resolution 50Hz data which is again sufficient to measure these FACs having a spatial extent of approximately 1000km. However, \citeA{Ritter2006} warn that the multi-spacecraft method of determining the FACs only respond well to FACs having scale sizes of more than 4 times the spacecraft separation. Although the sampling cadence can go as high as 20ms, the magnetic field data is generally low pass filtered to a cadence of 20s. This is still sufficient in terms of the temporal resolution required to capture such sharp changes in the FACs due to the cusp-FTE interaction. 

Finally, this study presents a morphological similarity between a SWARM observation of a FAC pair between 21:08 UT and 21:10 UT on October 7, 2015. As seen from the red curves in panel (a) and panel (b) of figure \ref{fig:obs_fte1}, there exists a very good agreement in the morphology, however, the peak magnitudes of the FACs obtained from the simulation is approximately 40 time lower. The FACs due to cusp-FTE interaction obtained from the simulation is consistently seen to be close to $\pm$1$\mu Am^{-2}$.  The magnetosheath and the high altitude cusps operate in a collisionless regime. As such, the collisionless reconnection rate is expected to be much higher than what can be obtained within the MHD approximation. As a result, the FAC magnitudes are also expected to be higher accordingly. We note, however, that the similarity in morphology of the simulated ionospheric signatures of FTEs corresponds well with one particular observational event and as such, it remains to be determined whether this correlation can be generalised for other events in which FAC pairs are observed. Nevertheless, we are only beginning to understand the impact that transient phenomena such as flux transfer events may have on the ionospheric surface.

\appendix
\section{Magnetosphere Ionosphere Coupling in \texttt{\large{M\MakeLowercase{ag}PIE}}}\label{appendix1}
\subsection{General Methodology}
Field aligned currents (FACs) are the dominant carrier of information from the global magnetosphere to the ionosphere. It is however not feasible to completely simulate the magnetosphere and the ionosphere as parts of a single Global MHD simulation due to model limitations. Global MHD simulations, therefore, generally incorporate a two-way coupled magnetosphere-ionosphere domain where the dynamics of the two separate domains are governed by different mathematical formulations. The information of the changes in the MHD domain is communicated to the ionosphere with the help of FACs whereas the impact of an ionosphere in the physical MHD system is emulated in the simulation by applying its feedback over a typical internal MHD boundary represented by a spherical surface, generally placed anywhere between 2.5 - 5 $R_E$ \cite{Daum_2008, Sun_2019}. In this section, we describe our integration of a flexible, general purpose, two-way coupled Magnetosphere-Ionosphere module within the solar wind-magnetosphere interaction model developed by \citeA{Paul_2022} using the resistive-MHD framework of the PLUTO code. The model has been named \texttt{MagPIE} (\textit{\textbf{Mag}netosphere of \textbf{P}lanets with \textbf{I}onospheric \textbf{E}lectrostatics}). The general workflow  of the magnetosphere-ionosphere coupling is described as follows.

\begin{figure}
    \centering
    \includegraphics[width=0.4\textwidth]{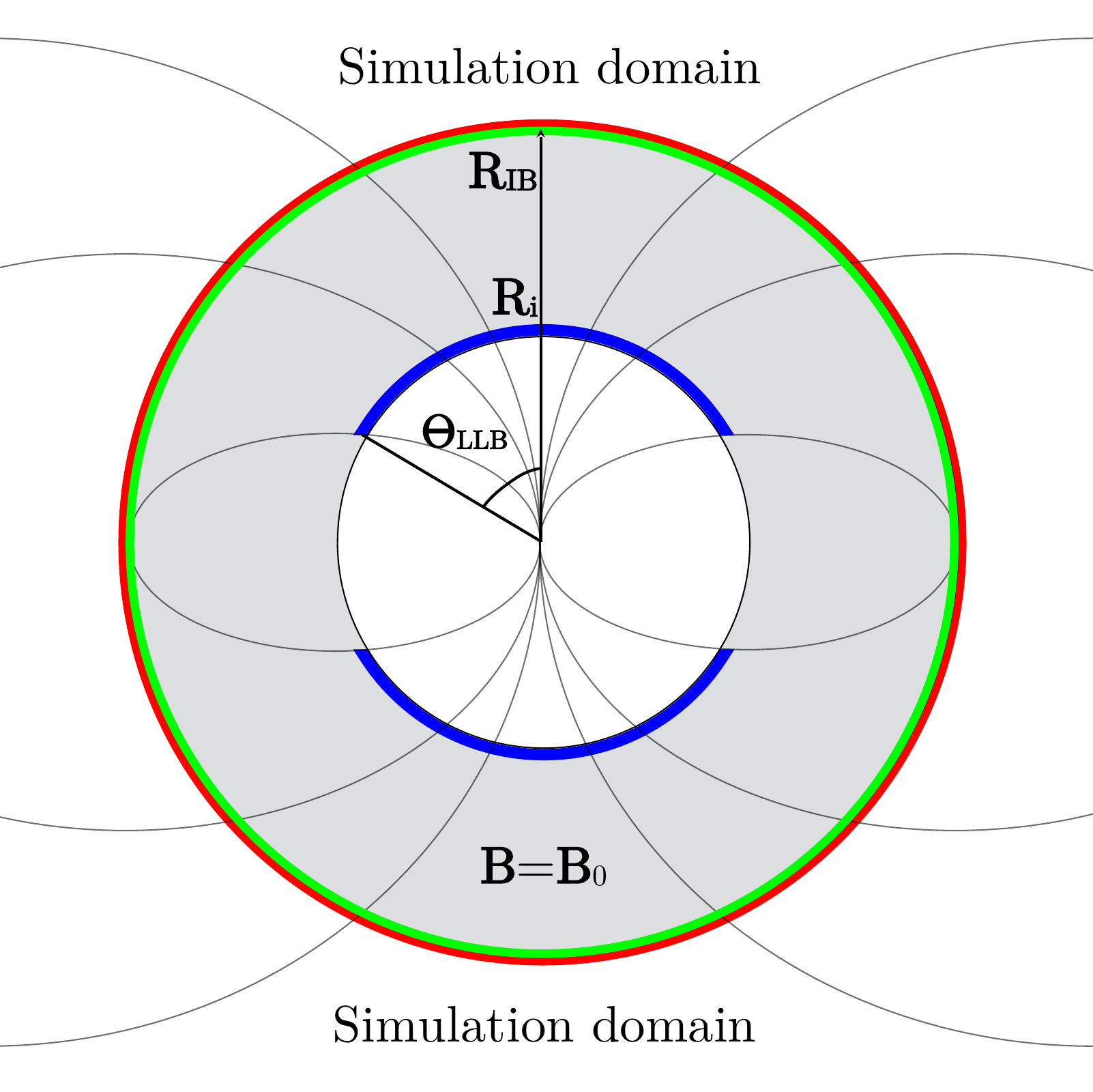}
    \caption{A schematic showing certain elements of the global MHD domain and the ionospheric model. The outer red circle denotes the layer responsible for depositing the FACs to the ionospheric module. The last closed field line from the global MHD domain reaches the ionosphere at a colatitude denoted by $\Theta_{LLB}$. The green circle just below this layer denotes the grid shell where the output of the ionosphere module are fed back to the global MHD domain. The bold blue arcs represent the colatitude extents of the ionospheric model. The magnetic field in the region between the internal boundary and the ionospheric shell is assumed to be purely dipolar having a field strength conforming to the planetary dipole (B$_0$).}
    \label{fig:MI_schematic}
\end{figure}

The northern and the southern hemispheres in this formulation are treated separately. From the global magnetospheric domain, we first calculate the field aligned currents (FACs) as $J_\parallel$ = $\mathbf{J}\cdot\mathbf{B}/|B|$ for any arbitrary inclination of the planetary dipole. We then perform a coordinate transform to obtain the FACs in a frame where the planetary field is vertical by an application of the rotation matrix. The ionosphere is considered to be a thin shell of infinitesimal thickness and the conductances are incorporated in a height integrated form with the height integrated Pedersen conductivity denoted as $\Sigma_P$ and Hall conductance denoted by $\Sigma_H$. The field aligned currents obtained in the aforementioned new coordinate frame where the dipole is vertical, is mapped down to the ionosphere to a height of 1.05 $R_E$ from the origin (0.05 $R_E$ from the surface of the Earth). For convenience, henceforth, we use spherical polar coordinates ($R,\theta,\varphi$). The mapping procedure is carried out as follows.

For the northern hemisphere, consider an infinitesimal flux tube that intersects a spherical surface of radius R= $R_{IB}$, where $R_{IB}$ is the radius of the MHD internal boundary. The flux tube will also intersect the ionospheric surface at a radius of R= $R_i$ and the conservation of charges between these two intersections requires that:

\begin{equation}
  \label{eq:conserv_charge}
  J_{\parallel}(R_{IB},\theta_{IB},\varphi_{IB}) R^2_{IB}{\rm d}\Omega_{IB}
  \cos\varepsilon_{IB} = \ J_{\parallel}(R_i,\theta_i,\varphi_i) R^2_i{\rm d}\Omega_i
  \cos\varepsilon_i
\end{equation}

Where the subscript `\textit{IB}' denotes quantities at the MHD internal boundary and the subscript `\textit{i}' denotes quantities at the ionospheric spherical shell. $\varepsilon$ is the angle between the planetary magnetic field and the normal to the local spherical surface. The colatitude ($\theta$) dependence of the quantity $\cos\varepsilon$ is given for a vertical dipole by \cite{Merkin_lyon_2010}:
\begin{equation}
\label{eq:NH_cose}
\cos\varepsilon = \frac{-2\cos\theta}{\sqrt{(1+3\cos^2\theta)}}
\end{equation}
 As the planetary magnetic field is vertical and dipolar, the magnetic field lines also satisfy the relation:
 \begin{equation}
 \label{eq:dipole_lines}
 \sin^2\theta_i = \left(\frac{R_i}{R_{IB}}\right)\sin^2\theta_{IB}   
 \end{equation}
 This further means that given a radius of the internal boundary, there exists a limit in the ionospheric colatitude to which the equatorial closed field line of the IB maps to. This limits the colatitude extent of the ionospheric grid to the bold blue arcs as denoted in figure \ref{fig:MI_schematic}. We denote this angle as $\Theta_{LLB}$ and its value is given by:
 \begin{equation}\label{eq:thetaLLB}
     \Theta_{LLB} = \sin^{-1}\left(\sqrt{\frac{R_i}{R_{IB}}}\right)
 \end{equation}. 
 
 Now, using equations \eqref{eq:NH_cose} and \eqref{eq:dipole_lines}, equation \eqref{eq:conserv_charge} can easily be expressed as:  

\begin{equation}
  \label{eq:jpar_map}
  J_{\parallel}(R_i,\theta_i,\varphi_i) = \left(\frac{R_{IB}}{R_i}\right)^3
  \sqrt{\frac{1+3\cos^2\theta_i}{1+3\cos^2\theta_{IB}}} \left[
  J_{\parallel}(R_{IB},\theta_{IB},\varphi_{IB})\right]\, .
\end{equation}

Equation \eqref{eq:jpar_map} is then used to map the FACs from the global-MHD internal boundary to the ionsopheric shell. 

The potential equation in the ionosphere (infinitesimal spherical shell of radius $R_i$) as given by \citeA{Goodman1995} can be expressed as

\begin{align}
  R_i^2 j_n(R_i,\theta,\varphi) &=
  \frac{\Sigma_P}{C}\partial_{\theta\theta}^2 \psi +
  \frac{1}{\sin^2\theta}\left(\Sigma_P +
  \frac{\Sigma_H}{\Sigma_0}\frac{\Sigma_H\sin^2\varepsilon_i}{C}\right) \partial_{\varphi\varphi}^2
  \psi \nonumber \\
  &+ \left[\cot\theta\frac{\Sigma_P}{C}
     + \partial_\theta\left(\frac{\Sigma_P}{C}\right) +
    \frac{1}{\sin\theta}\partial_\varphi\left(\frac{\Sigma_H\cos\varepsilon_i}{C}\right)
    \right]\partial_\theta\psi \nonumber \\
  &+\left[\frac{1}{\sin^2\theta}\partial_\varphi\left(\Sigma_P+\frac{\Sigma_H}{\Sigma_0}\frac{\Sigma_H\sin^2\varepsilon_i}{C}\right)
    -
    \frac{1}{\sin\theta}\partial_\theta\left(\frac{\Sigma_H\cos\varepsilon_i}{C}\right)\right]\partial_\varphi\psi
    \, ,
  \label{eq:pot_eq}
\end{align}
where $\psi$ is the ionospheric potential, $C=\cos^2\varepsilon_i + (\Sigma_P/\Sigma_0)\sin^2\varepsilon_i$, $j_n$ is the current density normal to the ionospheric shell at radius $R_i$, and $\varepsilon_i$ is the angle between the magnetic field and the normal to ionospheric shell boundary and is related to $J_{\parallel}$ as:
\begin{equation}
  \label{eq:jpar_jn}
  j_n(R_i,\theta,\varphi) = J_{\parallel}(R_i,\theta,\varphi)\cos\varepsilon_i
\end{equation}

$\Sigma_0$ in equation \eqref{eq:pot_eq} is the height integrated conductance along the magnetic field lines, $\Sigma_P$ is the Pedersen conductance and $\Sigma_H$ is the Hall conductance. The conductance along the field lines are generally considered to be much greater then the Pedersen and Hall conductances ($\Sigma_0 \gg \Sigma_P,\Sigma_H$), which simplifies the first and third terms in the RHS of Equation \eqref{eq:pot_eq} and leads to $C=\cos^2\varepsilon_i$. Under this assumption, we thus obtain

\begin{align}
  R_i^2 j_n(R_i,\theta,\varphi) &= \nonumber
  \frac{\Sigma_P}{\cos^2\varepsilon_i}\partial_{\theta\theta}^2 \psi +
  \frac{\Sigma_P}{\sin^2\theta} \partial_{\varphi\varphi}^2
  \psi \\ \nonumber
  &+ \left[\cot\theta\frac{\Sigma_P}{\cos^2\varepsilon_i}
    + \partial_\theta\left(\frac{\Sigma_P}{\cos^2\varepsilon_i}\right) +
    \frac{1}{\sin\theta}\partial_\varphi\left(\frac{\Sigma_H}{\cos\varepsilon_i}\right)
    \right]\partial_\theta\psi \nonumber \\ 
  &+\left[\frac{1}{\sin^2\theta}\partial_\varphi\left(\Sigma_P\right)
    -
    \frac{1}{\sin\theta}\partial_\theta\left(\frac{\Sigma_H}{\cos\varepsilon_i}\right)\right]\partial_\varphi\psi
    \, .
  \label{eq:pot_eq_reduced}
\end{align}

Equation \eqref{eq:pot_eq_reduced} is then solved numerically using a preconditioned generalized minimal residual method (GMRES) method. Dirichlet conditions are applied to the low-latitude ionospheric boundary denoted by $\Theta_{LLB}$ in figure \ref{fig:MI_schematic}. Following \citeA{Merkin_lyon_2010}, a Neumann boundary condition is applied at the poles which, in our current implementation, translates to 
\begin{equation}
\label{eq:North_pole}  
  \psi_{(0,j)} = \psi_{(2,j)} + \frac{j_{n (1,j)}}{\Sigma_{P0}}(\Delta\theta_0)^2
\end{equation}
where the subscript `0' represents the first row of the discretized ($\theta , \varphi$) grid (assuming $\theta$ is constant along a row and $\varphi$ is constant along a column), subscript `1' the second, and so on.

Once a solution $\psi(R_i,\theta_i,\varphi_i)$ for the ionospheric potential is computed, we calculate the spatial derivatives $\partial_{\theta_i}  \psi\left(R_i,\theta_i,\varphi_i\right)$ and $\partial_{\varphi_i}\psi\left(R_i,\theta_i,\varphi_i\right)$ on the regular spherical grid in the ionosphere and then propagate these quantities to the MHD internal boundary. At the spherical surface of the internal boundary, considering $E=-\nabla \psi$, the three components of the electric fields can be represented as:
\begin{align}
  \label{eq:E_phi}
  E_{\varphi_{IB}} &= - \frac{1}{R_{IB}\sin\theta_{IB}} \partial_{\varphi_{IB}}
  \psi\left(R_{IB},\theta_{IB},\varphi_{IB}\right) =  - \frac{1}{R_{IB}\sin\theta_{IB}} \partial_{\varphi_i}
  \psi\left(R_i,\theta_i,\varphi_i\right)\,  ,\\
  \label{eq:E_theta}
  E_{\theta_{IB}} &= - \frac{1}{R_{IB}} \partial_{\theta_{IB}}
  \psi\left(R_{IB},\theta_{IB},\varphi_{IB}\right) =  - \frac{1}{R_{IB}}
  \left( \partial_{\theta_{IB}} \theta_i \right) \partial_{\theta_i}
  \psi\left(R_i,\theta_i,\varphi_i\right) \nonumber \\ &= -\frac{1}{R_{IB}}
  \left( \frac{1}{\sqrt{1-\frac{R_i}{R_{IB}}\sin^2\theta_{IB}}}\sqrt{\frac{R_i}{R_{IB}}}\cos\theta_{IB} \right) \partial_{\theta_i}
  \psi\left(R_i,\theta_i,\varphi_i\right) \nonumber \\ 
&= -\frac{1}{R_{IB}}\sqrt{\frac{R_i}{R_{IB}}}\frac{\cos\theta_{IB}}{\cos\theta_i}
\partial_{\theta_i}
  \psi\left(R_i,\theta_i,\varphi_i\right)\, ,\\
  \label{eq:E_r}
  E_{R_{IB}} &= - \partial_{R_{IB}}
  \psi\left(R_{IB},\theta_{IB},\varphi_{IB}\right) =  -  \left( \partial_{R_{IB}} \theta_i \right) \partial_{\theta_i}
  \psi\left(R_i,\theta_i,\varphi_i\right) \nonumber \\ &= \frac{1}{2R_{IB}}
  \frac{1}{\cos\theta_i}\sqrt{\frac{R_i}{R_{IB}}}\sin\theta_{IB} \partial_{\theta_i}
  \psi\left(R_i,\theta_i,\varphi_i\right) \nonumber \\ 
&= \frac{\tan\theta_i}{2R_{IB}}\partial_{\theta_i}
  \psi\left(R_i,\theta_i,\varphi_i\right)\, .
\end{align}

These fields are then converted back into the global-MHD frame by the application of an inverse rotation matrix and finally, the velocity at the surface of the internal boundary is specified in the form of a drift given by the relation:

\begin{align}
  \label{eq:v_BC}
  {\bf v}_{\psi} = \frac{{\bf E} \times \mathbf{B}}{B^2}\, ,
\end{align}
where `\textbf{B}' is the local magnetic field. Additionally, due to the timescales involved in the current study, the corotation velocity of the ionospheric plasma is expected to be negligible and therefore has not been considered in this implementation, however, it can be easily added when required. 

\subsection{Validation of the Coupled Ionospheric Module}
In this section, we establish the validity of our magnetosphere-ionosphere coupling by a qualitative analysis of the global-MHD input to the ionospheric module and the module output. We omit any exact quantitative comparisons of the model due to the fact that owing to differences such as grid geometry, resolution, and numerical approaches used in different global-MHD codes, the principal metrics such as the input $J_{\parallel}$ and the output $\psi$ can vary up to an order of magnitude in their values even for very similar solar wind driving conditions \cite{Honkonen_2013, Gordeev_2015, eggington2021_thesis}. The solar wind conditions used for validation is the same as described in the section \ref{sec:Initial_and_boundary}. We note here that before integrating the ionospheric solver into PLUTO, we have also validated the solver as a standalone piece of code against the test problems described in \citeA{Merkin_lyon_2010}.

\begin{figure*}
    \centering
    \includegraphics[width=0.8\textwidth]{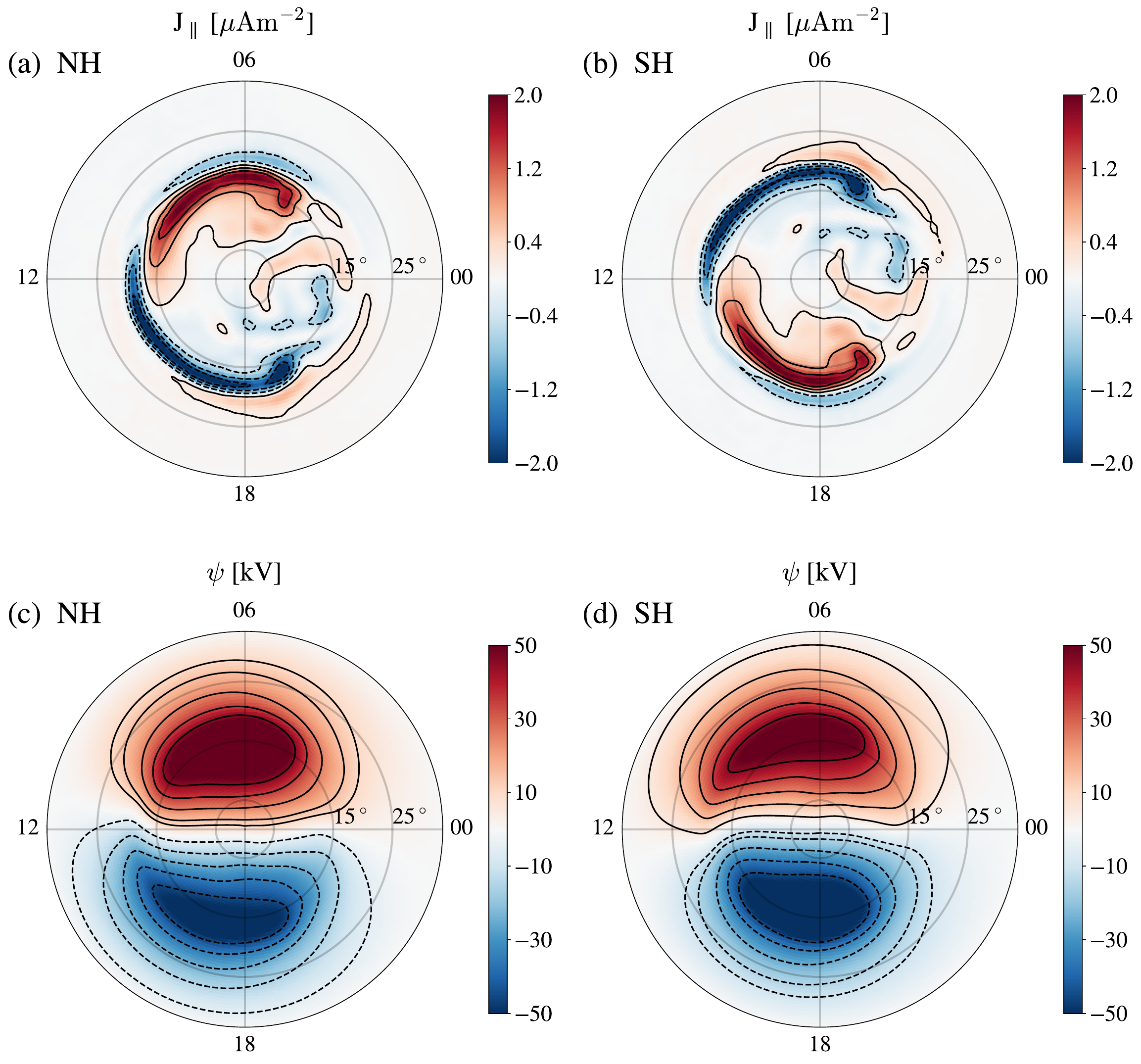}
    \caption{A plot of the inputs and outputs of the ionospheric potential solver. The left column (panels (a) and (c)) highlights the northern hemisphere (`NH') and the right column (panels (b) and (d)) highlights the southern hemisphere (`SH'). The panels (a) and (b) correspond to the FACs (J$_\parallel$) calculated on the ionospheric surface for the northern and the southern hemispheres respectively whereas the panels (c) and (d) show the corresponding ionospheric potential ($\psi$) obtained. The contours in all the subplots are for a better visualisation of the morphology.}
    \label{fig:MI_jpar_psi}
\end{figure*}

We first look into the FAC pattern produced in the ionosphere by the global MHD model towards the end of the simulation when the IMF has been constantly southward for $\sim$ 1 hour. Panels (a) and (b) in figure \ref{fig:MI_jpar_psi} show the $J_{\parallel}$ produced in the northern hemisphere (NH) and the southern hemisphere (SH) on the ionospheric shell. The polar plots used in these representations (and all representations henceforth) have the magnetic local time (MLT) in the azimuthal axis. An MLT of 00 represents the midnight sector, whereas an MLT of 12 stands for the noon sector. Similarly, MLT vales of 06 and 18 represent the dawn and dusk sectors respectively. The radial extent of the plot represents the colatitude grids starting from 0$^\circ$ at the North and south poles and ending at the low latitude boundary ($\Theta_{LLB}$) which has a value of 33.2$^\circ$ following equation \eqref{eq:thetaLLB} with $R_i = 1.05 R_E$ and $R_{IB} = 3.5 R_E$.

The FAC pattern shows strong Region-1 (R1) current arcs with peaks around $\sim$ 18$^\circ$ colatitude with the comparatively fainter Region-2 (R2) currents equatorward, peaking at $\sim$ 21$^\circ$ colatitude.
As can be clearly seen from the panels (a) and (b) of figure \ref{fig:MI_jpar_psi}, the structure of the FACs obtained from our simulation is morphologically similar to the statistical pattern of large scale FACs seen above the auroral oval consisting of currents into and away from the ionosphere \cite{Iijima_1976, Anderson_2008, Koskinen_2011}. The magnitudes of the current peaks are also well in agreement with recent observations during southward IMF conditions \cite{Xiong2020, Pedersen_2021}. Compared to the NH, the FACs in the SH exhibit a sign reversal in the dawn and dusk sector due to the reversed sense of the dipolar magnetic field lines in the two hemispheres. 

Panels (c) and (d) of figure \ref{fig:MI_jpar_psi} show the ionospheric potential of calculated by the ionospheric solver. Due to the extended southward IMF period, the cross polar cap potential peaks at $\sim$ 120kV for both the hemispheres. For better visualisation, the color bars in the panels (c) and (d) has been capped at $\pm$50kV. The contours show the spatial distribution pattern of the potential, exhibiting the typical two lobe structure expected for a southward IMF condition. They also highlight that the obtained solution has a smooth transition across the coordinate singularity at the pole, which reveals that the pole boundary condition implementation as given in equation \eqref{eq:North_pole} is working as expected. As seen from the two plots, the potential remains the same for the dawn and dusk sectors in both the hemispheres. The slight asymmetry with one of the potential lobes being larger than the other is due to the finite $B_y$ component of the incoming solar wind, and the pattern is consistent with the statistical pattern of the potential for positive IMF $B_y$ \cite{Chisham2007,Holappa_2021}.

\begin{figure}
    \centering
    \includegraphics[width=0.8\textwidth]{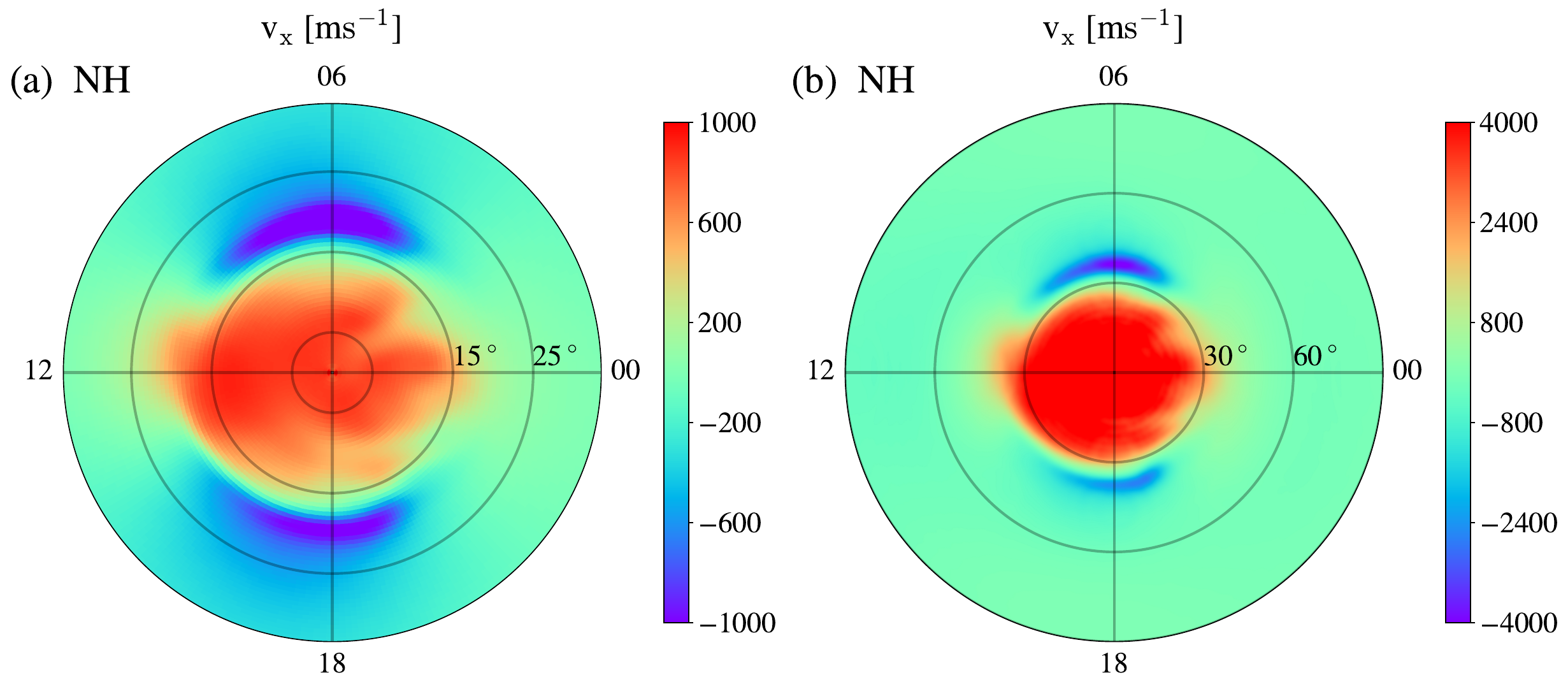}
    \caption{Panel (a) shows the drift velocity calculated on the ionospheric grid using equation \eqref{eq:v_BC} at an ionospheric height of 1.05 $R_E$ using the electric field obtained from the output of the potential solver. The theta grids in panel (a) are limited to the ionospheric grid colatitudes, i.e 33.2 $^\circ$. Panel (b) shows the velocity that is imposed on the internal boundary of the MHD grid at 3.5 $R_E$. The theta grid in panel (b) covers the entire northern hemisphere of the MHD internal boundary i.e $\theta =[0 ,90 ^\circ]$.}
    \label{fig:MI_vx}
\end{figure}

We now investigate the coupling of the ionospheric model to the global-MHD model in terms of the final task assigned to the module, i.e; the drift velocities given by equation \eqref{eq:v_BC} which are deposited on the global-MHD internal boundary. As a test, we first calculate the drift velocity on the ionospheric shell at 1.05 $R_E$ to review the expected pattern of the velocities obtained from the ionospheric potential. The x-component of the resulting velocity field is plotted in panel (a) of figure \ref{fig:MI_vx} that highlights the sunward or the antisunward velocities. As can be clearly seen, the solution exhibits strong anti-sunward flows near the poles up to $\sim$ 18$^\circ$ colatitude with return sunward flow at higher colatitudes (towards the equator). This pattern is in agreement with the ionospheric convection pattern proposed by \citeA{Dungey_1961} for a reconnecting magnetosphere \cite{Koskinen_2011}. Panel (b) of figure \ref{fig:MI_vx} shows the $v_x$ directly sampled from the MHD domain. The ionospheric grid, when mapped out to the IB along the dipole field lines spans the entire northern hemisphere and therefore, the $\theta$ grids in panel (b) span up to 90$^\circ$ colatitude. It is seen that the profile of velocities deposited on the IB surface is consistent with panel (a) with the velocity reversal from anti-sunward to sunward motion mapped to approximately 32$^\circ$ colatitudes. We also note that the velocities imposed at the IB height has a greater magnitude than the values calculated at the ionospheric shell. We further verify that the southern hemisphere is also consistent with the expected results and the transition of the separately obtained solutions across the equator is smooth. Finally, we conclude from a careful observation of this set of outputs that the MI module is working as expected and the desired coupling between the global-MHD model and the ionospheric model is being achieved to a reasonable degree.

 \subsection{FACs and Potential during FTE-Ionosphere Interaction}
\begin{figure*}
    \centering
    \includegraphics[width=1.1\textwidth]{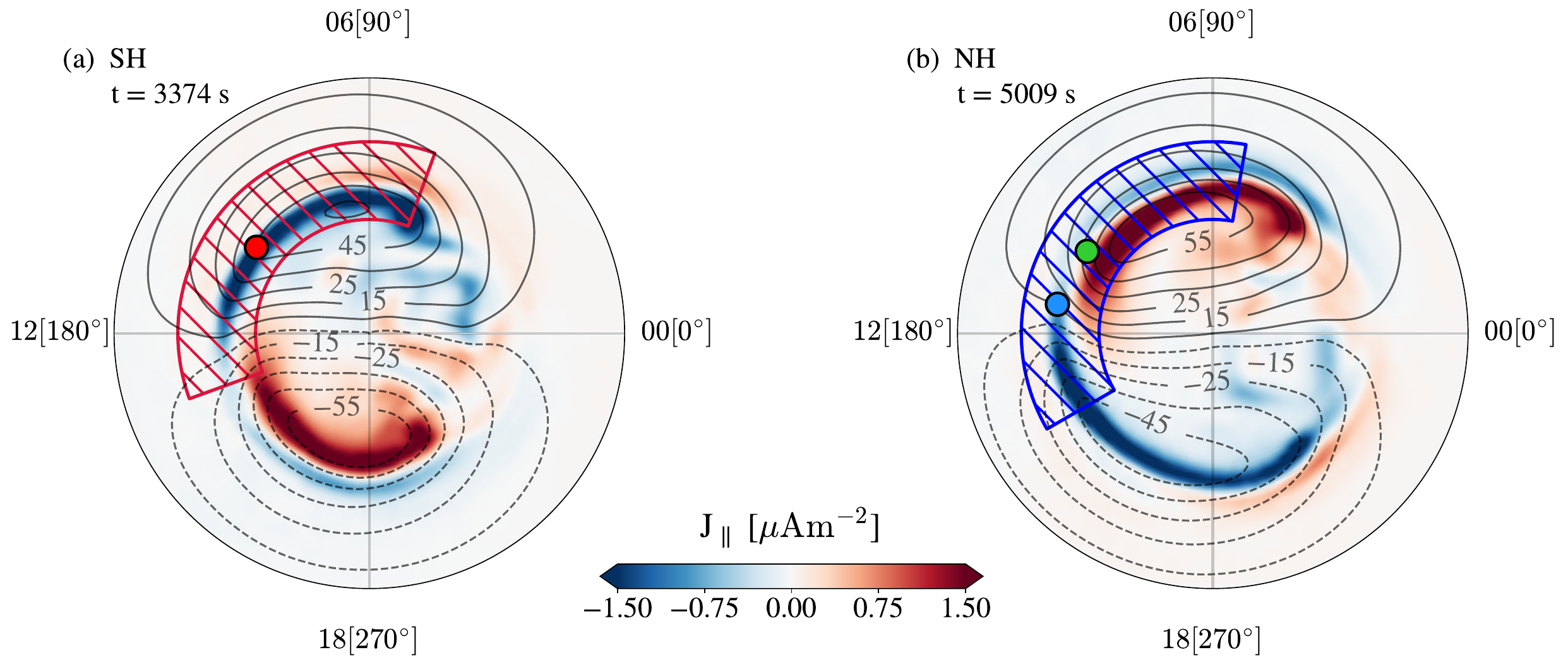}
    \caption{Plots showcasing various regions of the ionospheric module used in describing the results presented in section \ref{sec:ionospheric_signals_FTEs}. The left plot corresponds to the southern hemisphere whereas the right plot represents the northern hemisphere. The background color in both the panels represent the large scale R1 and R2 current systems obtained from the ionospheric module in the respective hemispheres. The solid and the dotted contours highlight the morphology of the positive and the negative lobes of the corresponding polar cap potential. The red and blue hatched regions highlight the ionospheric subsections featured in figures \ref{fig:fte1_jpar_pot} and \ref{fig:fte2_jpar_pot} respectively whereas the red, green and blue scatter points correspond to the satellite locations in the results highlighted in section \ref{sec:sat_signatures}. }
    \label{fig:fte_location_shade}
\end{figure*}

  The presence of a southward IMF results in an enhanced magnetosphere ionosphere coupling, and therefore, the large scale FACs obtained from \texttt{MagPIE} are expected to be quite strong in magnitude. Here we present the morphology of the large scale currents and potential during the evolution of FTE-1 and FTE-2 for reference. The background pseudocolor in panels (a) and (b) of figure \ref{fig:fte_location_shade} shows the morphology of the large scale FACs on the southern and the northern ionospheric surfaces at two different times, i.e., $t \sim 3374s$ and $t \sim 5009s$ respectively (approximately in between the duration of the ionospheric response produced by these FTEs as discussed in section \ref{sec:ionospheric_signals_FTEs}). Owing to the southward IMF, the ionospheric currents exhibit the typical structure of Region-1 (R1) current arcs indicated by the strong poleward FACs. The Region-2 (R2) current system is the weaker equatorward FAC arcs that are located adjacent to the R1 currents. The solid and dashed oval lines in panels (a) and (b) are equipotential contours that outline the polar cap potential structure. The peak FAC values in panel (a) reaches 2.7$\mu A m^{-2}$ and that in panel (b) reaches 2.8$\mu A m^{-2}$. The corresponding cross polar cap potentials have values of 117kV and 119kV respectively indicating a strong magnetosphere-ionosphere coupling. The panels (a) and (b) of figure \ref{fig:fte_location_shade} also serve as a general reference for the large scale FAC and potential morphologies during the impact of FTE-1 and FTE-2 on the ionosphere as discussed in section \ref{sec:ionospheric_signals_FTEs}.

\subsection{A Note on the interpolation of $J_{\parallel}$ to a regular ionospheric grid}
The mapping of the quantity $J_{\parallel}$ from the MHD domain to the ionospheric shell is done following the dipolar magnetic field lines from the internal boundary. As such, the regularly spaced cartesian grid cells that constitute the surface of the internal boundary, form a set of irregularly spaced points when mapped back to the ionospheric surface. To decrease the complexity of the potential solver, we first interpolate these irregularly spaced set of points into a regularly spaced uniform spherical grid in ($\theta,\phi$). The interpolation method used here is known as Thin-Plate-Splines(TPS) which has gained widespread adoption due to its robustness. The complexity of the method, however, is such that it requires a looping through all the data points on the surface of the IB to determine the value of each cell on the ionospheric grid. At high resolutions, the number of cells constituting the spherical surface of the IB could easily reach values of the order $10^4$. This means that even for a modest ionospheric resolution of the order of $10^2 \times 10^2$, the number of iterations could easily fall within a range of $10^8 - 10^9$ which severely slows down the ionospheric solver. To tackle this issue, we selectively sample the set of points that deposit their value of $J_{\parallel}$ into the ionospheric grid following a procedure described below.

First, all points within a 2$^\circ$ colatitude from the two poles are included. Then, from a low resolution run, the colatitude variation of the transition region between the R1 and R2 current system on the IB surface is determined. This colatitude will have a longitudinal dependence and we represent it as $\theta_{cusp} (\varphi)$. For the northern hemisphere, we then consider a surface that lies within $\theta (\varphi) = [0,\; \theta_{cusp} (\varphi) + 15 ^\circ]$ bounds and select half of the points from this surface to deposit their values of $J_{\parallel}$ on the ionospheric grid. An analogous treatment is performed for the southern hemisphere as well. This exercise ensures that most of the grid cells that deposit their value of $J_{\parallel}$ to the ionospheric grid contribute to the large scale R1 and the R2 current systems. For all the points that remain outside these two partial hemispheric sections, we select 10\% of the grid cells at random and deposits their $J_{\parallel}$ on the ionospheric grid thereby accommodating regions from the entire IB surface. We have ensured that such a selective sampling only slightly affects the maximum magnitude of $J_{\parallel}$ interpolated in the ionospheric grid and has a negligible effect on the temporal and spatial profiles of the FACs.


%
%



\bibliography{MS_AP}

%
%
%
%
%

\section{Open Research}
The astrophysical gasdynamics code PLUTO that is used for the numerical simulation is open source and can be downloaded from \url{http://plutocode.ph.unito.it/} \cite{pluto-4.3}. Figure 9(a) uses a part of the SWARM data first published in \citeA{Dong_2023}. The SWARM data is publicly available at  \url{https://swarm-diss.eo.esa.int/} and the FAC data from two spacecrafts, Swarm A and C, named \texttt{SW\_OPER\_FAC\_TMS\_2F} has been used.

\acknowledgments
The authors thank the anonymous reviewers whose constructive comments and suggestions have significantly improved the manuscript. A.P. is a research scholar at IIT Indore and would like to acknowledge the support provided by the institute. A.S. acknowledges funding from the French Programme National Soleil-Terre (PNST) and Programme National de Planétologie (PNP). B.V. is thankful for the support provided by the RESPOND grant (ISRO/RES/2/436/21-22) given by Indian Space Research Organization to support this research. He also acknowledges the support for leading the Max Planck Partner Group at IIT Indore provided by the Max Planck Society. All computations presented in this work have been carried out using the facilities provided at IIT Indore, the Max Planck Institute for Astronomy Cluster: VERA which is a part of the Max Planck Computing and data Facility (MPCDF), and the CEA DAP/AIM clusters ALFV\'EN and ANAIS founded by DIM ACAV +.

\end{document}